\documentclass[prd,aps,reprint,amsmath,amssymb,twocolumn,nofootinbib,floatfix]{revtex4-2}

\usepackage{wasysym,graphicx}
\usepackage{epstopdf}
\usepackage{alphalph}

\usepackage[colorlinks,linkcolor=blue]{hyperref}
\usepackage[normalem]{ulem}

\usepackage{url}

\usepackage{academicons}
\usepackage[caption=false]{subfig}
\usepackage{ragged2e} % for the \justifying macro
\DeclareCaptionJustification{justified}{\justifying}
\def\al{\alpha}
\def\be{\beta}
\def\ga{\gamma}
\def\de{\delta}
\def\ep{\epsilon}
\def\ze{\zeta}
\def\et{\eta}
\def\th{\theta}
\def\ka{\kappa}
\def\la{\lambda}
\def\rh{\rho}

\def\ta{\tau}

\def\ph{\phi}

\def\Ga{\Gamma}
\def\De{\Delta}

\def\La{\Lambda}
\def\Si{\Sigma}

\def\Om{\Omega}

\def\mn{{\mu\nu}}
\def\ab{{\al\be}}

\def\tV{{\tilde V}}
\def\t#1{{\tilde #1}}

\def\prt{\partial}
\def\pt#1{\phantom{#1}}
\def\cL{{\cal L}}
\def\rf#1{(\ref{#1})}

\def\af#1{a_{f#1}}%horizon

\def\cf#1{c_{f#1}}%asymptotic

\def\cN#1{c_{N#1}}

\newcommand{\beq}{\begin{equation}}
\newcommand{\eeq}{\end{equation}}
\newcommand{\bal}{\begin{aligned}}
\newcommand{\eal}{\end{aligned}}

\begin{document}

\title{Bumblebee gravity: spherically-symmetric solutions away from the potential minimum}

\author{Quentin G.\ Bailey}
 \email{baileyq@erau.edu}
\author{Hailey S.\ Murray}
\author{Dario T.\ Walter-Cardona}
\affiliation{
Department of Physics and Astronomy
Embry-Riddle Aeronautical University\\
3700 Willow Creek Road, Prescott, Arizona, USA
}

\date{\today}

\begin{abstract}
In this work, 
we study a vector model of spontaneous spacetime-symmetry breaking coupled to gravity:
the bumblebee model.
The primary focus is on static spherically symmetric solutions.
Complementing previous work on black hole solutions, 
we study the effects on the solutions when the vector field does not lie at the minimum of its potential.
We first investigate the flat spacetime limit, 
which can be viewed as a modified electrostatic model with a nonlinear interaction term.
We study the stability of classical solutions generally and in the spherically-symmetric case. 
We also find that certain potentials, based on hypergeometric functions, 
yield a Hamiltonian bounded from below for the case of fixed spatial vector magnitude. %change here
With gravity, 
we solve for the spherically-symmetric metric and vector field, 
for a variety of choices of the potential energy functions, 
including ones beyond the quadratic potential like the hypergeometric potentials.
Special case exact solutions are obtained showing Schwarszchild-Anti de Sitter and Reissner-Nordstrom spacetimes. 
We employ horizon and asymptotic analytical expansions along with numerical solutions to explore the general case with the vector field away from the potential minimum. 
We discover interesting features of these solutions including naked singularities, repulsive gravity,
and rapidly varying gravitational field near the source.
Finally, 
we discuss observational constraints on these spacetimes using the resulting orbital behavior.
\end{abstract}

\maketitle

\section{Introduction}

Since shortly after the inception of General Relativity (GR), physicists proposed alternative models.
Those that survived experimental constraint are typically based on a fully dynamical model with an action and one or more extra tensor fields in spacetime with their own dynamics, 
the scalar field being the most notable \cite{Brans:1961sx}.
The rise and fall of such proposals has been discussed at length elsewhere \cite{Will:2018bme}.
Aside from the usual Einstein-Maxwell theory, 
in the 1970s, 
a vector field model was proposed which includes Einstein-Maxwell terms  but adds ghost kinetic terms for the vector field to the action \cite{wn72,hn73}.

In the late 1980s and 1990s it became of interest to reexamine the idea of extra non-scalar fields when it
was found that the ground state of string field theory may give rise to nonzero vacuum expectation values of tensor fields, 
thereby breaking the Lorentz symmetry of the vacuum \cite{ksstring89}.
Of central importance is the idea of a potential energy function for these fields that has a minimum for nonzero values for the tensor field, 
similar to the spontaneous-symmetry breaking of internal symmetries in the Standard Model of particle physics \cite{Weinberg67}.
The simplest case to consider is a vector field, 
on which we focus our attention in this work.

It has become clear of late
that black holes are a fruitful testing ground for new physics beyond GR and the Standard Model
\cite{LIGOScientific:2016lio,Bambi:2017iyh,Berti:2018cxi,LIGOScientific:2019fpa,Isi21}.
First and foremost, 
gravitational wave detection has allowed for the direct study of propagation and polarization effects of modified models going beyond GR \cite{km16,yunes16,Abbott_2017,Abbott_2019,Isi_2017,Hagihara_2019}. 
Also black hole imaging
has now become a measurable phenomenon
and the detailed behavior of light rays around a black hole is of interest \cite{Akiyama19}.

Since the 1990s, 
there has been substantial work on experimental and observational searches for hypothetical spacetime-symmetry breaking \cite{datatables,Safronova:2017xyt}, 
motivated by theoretical scenarios in which it could arise \cite{Addazi:2021xuf,Liberati13,ck98,k04,gp99}.
While no statistically-significant signal for a breakdown of Lorentz, CPT, or diffeomorphism symmetry has yet been found, 
the parameter space of possible effects remains large
\cite{km09,km13,bkx15,km16,kl21}. 
Studies of the effects of spacetime-symmetry breaking on black holes have been published in the last decades.\footnote{The literature is extensive, examples include Refs.\ \cite{Bertolami:2005bh,Barausse:2013nwa,Garfinkle:2007bk,Jacobson:2010fat,casana18,Xu:2022frb}.}

A general effective field theory (EFT) framework for studying spacetime-symmetry breaking is now widely used \cite{ck97,ck98,k04}.
It includes a gravity sector with arbitrary types of symmetry breaking for gravity \cite{kl21}, 
for example, 
there is the widely studied Ricci coupling $\cL \sim s_\mn R^\mn$, 
controlled by the coefficients $s_\mn$
\cite{k04, bk06, b09, Muller:2007es, Shao:2014oha, Bourgoin:2016ynf, Shao:2017bgz,Xu:2019gua,Jesus:2019nwi,Bonder:2020fpn,Bonder:2021gjo,abn21,Nilsson:2022mzq,Reyes:2021cpx,Lambiase:2023zeo,Nilsson:2023szw,Bailey:2023lzy,ABN24,Reyes:2024ywe,Lafkih:2024qva}.
While general dynamical terms have been countenanced for coefficients like $s_\mn$ \cite{b21}, 
it is still challenging to study in a systematic way, 
and so one is often led to special toy models.

Our aim in this work is to examine a class of vector models of spontaneous spacetime symmetry breaking in the context of black hole-type solutions.
In particular, 
we are interested in the effects of the potential for the vector field $B_\mu$, $V=V(B_\al B^\al)$ on gravitational physics, 
especially when the field configuration 
is {\it not assumed} to be at its minimum.

The plan of study is the following.
In Sec.\ \ref{bumblebee gravity background},
we review the background on the vector model to be studied.
First, 
we explore the flat spacetime limit in Sec.\ \ref{flat spacetime study}, 
with specific solutions presented or characterized for various cases of the potential $V$.
In particular, 
we establish some generic Hamiltonian properties in Sec.\ \ref{field equations, Hamiltonian}.
The full field equations coupled to gravity are the subject of Sec.\ \ref{gravity-vector, part I} and Sec.\ \ref{gravity-vector part II}.
Section \ref{gravity-vector, part I} discusses some generalities including energy theorems,
a simple cosmological solution,
the Bekenstein ``no-hair" argument, 
and special cases with exact solutions.

Specific approaches to solving the field equations for spherical symmetry are discussed in Sec.\ \ref{gravity-vector part II}.
In Sec.\ \ref{analytic}, 
we use analytical expansions from the horizon and the asymptotically flat region combined with numerical simulations. 
To contrast, 
we consider also a nonanalytic expansion in the asymptotic region in Sec.\ \ref{non-analytic}.
Finally, 
we make some remarks on measuring the parameters of the model through orbital perturbations and summarize our results in Sec.\ \ref{discussion and summary}.
Our conventions include the spacetime metric signature $-+++$, Greek letters are spacetime indices, 
Latin letters are spatial indices, and covariant derivatives are $\nabla_\mu$. 
Natural units with $\hbar =c =1$ are used. 

\section{Bumblebee gravity background}
\label{bumblebee gravity background}

A vector model of spontaneous Lorentz violation with a potential driving the symmetry breaking and a Maxwell kinetic term was first proposed in 
Ref.\ \cite{ks89bb}, 
dubbed the ``bumblebee model."\footnote{See Ref.\ \cite{seifert09} for an explanation of the name ``bumblebee."}
Another vector model, 
dubbed ``Einstein-Aether theory", 
modifies the kinetic terms in the action, and specializes the potential to be a Lagrange multiplier
\cite{Jacobson:2000xp}.
A plethora of research on the bumblebee model with nonminimal couplings 
in the context of special spacetime geometries has occurred in the last decade \cite{Bonder:2015jra,casana18,Ovgun:2018xys,Maluf:2020kgf,Jesus:2020lsv,Liang:2022hxd,Xu:2022frb,Izmailov_2022,Filho:2022yrk,Lambiase:2023zeo,Xu:2023xqh,PottingATT,Singh:2024nvx,Mai:2024lgk,Chen:2024pys,Ji:2024aeg,Neves:2024ggn,Filho:2024hri,Liu:2024axg,Belchior:2025xam,EslamPanah:2025zcm} and prior \cite{bk05,Bertolami:2005bh,bk08,Bluhm:2008yt,Carroll:2009em}.\footnote{These references are not exhaustive; a quick search on INSPIRE for ``bumblebee" reveals over $170$ results.}
This also includes quantum issues, 
such as a meaningful quantization program with potentials of different forms \cite{Hernaski:2014jsa,Maluf:2015hda,Delhom:2022xfo,Lehum:2024wmk,Araujo:2024tiy}.

For this work we assume Riemann geometry with no torsion. The Lagrange density for a generic bumblebee model is
\beq
\bal
\cL &= \sqrt{-g} \Big[ 
\frac {1}{2\ka} (R + \xi_1 B_\al B^\al R + \xi_2 B_\al B_\be R^{\al\be} )\\
&
\pt{space}
-\frac {1}{4} B^\mn B_\mn - V - B_\mu J^\mu
\Big] + \cL_M,
\label{bbmodel}
\eal
\eeq
where $B_\mn=\prt_\mu B_\nu - \prt_\nu B_\mu$ is the field strength tensor for $B_\mu$, 
and $\cL_M$ is the matter Lagrange density which we leave unspecified.
We include a possible coupling of the $B_\mu$
field to a conserved current $J_\mu$, 
in the case that $B_\mu$ is interpreted as the electromagnetic potential.
Note that one can add extra ``ghost" kinetic terms for $B_\mu$ to this action, or others,
which we do not consider in this work.
The potential depends on the only scalar contraction available for
a vector field:
\beq
V = V (B_\al B^\al).
\label{potential}
\eeq
Spontaneous symmetry breaking of spacetime symmetry occurs when the potential has a minimum at a nonzero value of $B_\mu$;
thereby defining the vacuum of the associated Quantum Field Theory.
Thus, 
the spacetime symmetry can be broken if
\beq
\langle B_\mu \rangle = b_\mu \neq 0.
\label{vac}
\eeq
The presence of the potential by itself explicitly breaks the $U(1)$ usual gauge symmetry of Einstein-Maxwell theory.

The remaining terms with $\xi_1$ and $\xi_2$ are the nonminimal couplings to curvature.
Note that much of the literature has focused only on the $\xi_2$, 
neglecting possible generalized phenomena with both parameters are in play \cite{ABN24}.
These couplings are of the form of the $s_\mn$ and $u$ type couplings in the gravitationally-coupled Standard-Model Extension framework \cite{k04,bk06}.
Much work has been devoted to searches for symmetry breaking effects of these terms under different assumptions 
for the mechanism for spacetime symmetry breaking.
However, 
our aim in this work is to focus on the effects of the potential $V$ when it does not lie at its minimum, 
which has not been studied extensively in the literature.
For this purpose we assume, in this paper,
\beq
\xi_1 = 0 = \xi_2 \,\,({\rm assumption \,\, for \,\,this \,\,work}).
\label{xi0}
\eeq

Note that the condition \rf{xi0} implies that this work differs from other phenomenological approaches where the effects of $\sim \xi BBR$ are the focus.  
The latter approaches generate background ``coefficients" that can be measured like $s_\mn \sim \xi b_\mu b_\nu$, 
and their presence clearly indicates symmetry breaking, 
at least in weak-field gravity scenarios.
Nonetheless, 
our focus on asymptotically flat solutions will generate values of $B_\mu$ in the asymptotic region that can be identified as vector coefficients for Lorentz violation akin to the $b_\mu$ coefficients in the EFT framework \cite{Kostelecky:2000mm}.

\section{Flat spacetime study}
\label{flat spacetime study}

\subsection{Field equations, Hamiltonian, and potential energy}
\label{field equations, Hamiltonian}

The bumblebee model in the absence of gravity has the following field equations ($J^\mu =0$ for simplicity):
\beq
\prt_\mu B^\mn = 2V' B^\nu,
\label{flatBB}
\eeq
where $V=V(\et^\mn B_\mu B_\nu)$, 
and the prime denotes the derivative with respect to the argument.
Note that taking a partial derivative yields the following constraint
on the dynamical variables
\beq
\prt_\nu (2 V' B^\nu) =0.
\label{flatBBc}
\eeq
In the case of a massive vector field, 
$V'=\frac 12 \mu^2$ this yields $\prt_\mu B^\mu=0$, 
which can then be used to simplify the equations.
For more general nonlinear" $V'$ cases, 
this trick does not work in the same way, 
although the condition \rf{flatBBc}
still represents a constraint on the field components \cite{bk05}.

We consider here an arbitrary potential $V$ and extend previous works for this case.
Before solving the equation, 
we review the Hamiltonian evolution, 
as it provides useful insight for what follows.
For special choices of $V$, 
the classical Hamiltonian dynamics have been studied \cite{Bluhm:2008yt,Bonder:2015jra}, 
along with similar models \cite{Seifert:2018mmr,Seifert:2019kuz}.
We adopt the standard Dirac-Hamiltonian procedure 
for a constrained system \cite{Anderson:1951ta,dirac1964}
and make use of results in previous works \cite{abn21}.
In particular, 
we identify the (starting) four field components:
$(B_0, B_j)$, 
and the corresponding conjugate momenta $\Pi^\mu = \de L/\de {\dot A}_\mu$.
The Lagrangian in this limit is $L=-\int d^3x (\tfrac 14 B^\mn B_\mn +V)$.
We first obtain the usual primary constraint
and the equation for $\vec \Pi$:
\beq
\bal
\Pi^0 &\approx 0,
\\
\vec \Pi &= \dot {\vec B} - \vec \nabla B_0,
\label{primary&pivec}
\eal
\eeq
where the wavy equals sign denotes a ``weak" equality, 
or equality only when the constraints are imposed.
The base, 
or beginning Hamiltonian density is ${\cal H}_0=\pi^\mu \dot{B}_\mu - \cL =\frac 12 \vec \Pi^2 + \frac 14 B_{ij}B^{ij} + V + \vec \Pi \cdot \vec \nabla B_0 + \Pi^0 \dot{A}_0$.
The next step is to check the evolution of the primary constraint $\dot{\Pi}^0 = \{\Pi^0,H\}$.
The brackets here are Poisson brackets and explicit formulas can be found in standard references, 
for example, 
see Appendix B in Ref.\ \cite{abn21}, or the appendix in Ref.\ \cite{Seifert:2018mmr}.
Ensuring the constraint is preserved, 
$\dot{\Pi^0}\approx 0$, 
yields the secondary constraint (modified Gauss's law):
\beq
\dot{\Pi}_0 = \vec \nabla \cdot \vec \Pi + 2 V' B_0 \approx 0.
\label{SC}
\eeq

After using the Hamiltonian with the primary constraint added $+u \Pi^0 $, 
$\cal H_A$,
and ensuring that the secondary constraint is preserved, 
one can solve for the Lagrange multiplier $u$ in terms of functions of the fields $\Pi^\mu$ and $B_\mu$.
The final Hamiltonian density for the constrained system
is ${\cal H}_A$ with the formula for $u$ inserted:
\beq
\bal
{\cal H}&  =\frac 12 \vec \Pi^2 + \frac 14 B_{ij}B^{ij} + V + \vec \Pi \cdot \vec \nabla B_0 \\
&\pt{sp}+ \Pi^0 \left[ \frac {\vec \nabla \cdot (V' \vec B)-V'' 2 B_0 \vec B \cdot (\vec \Pi + \vec \nabla B_0) }{-2 V'' B_0^2 + V'} \right].
\label{Hfin}
\eal
\eeq
With equations \rf{primary&pivec} and \rf{SC}, 
the remaining Hamilton equations of motion are
\beq
\bal
{\dot B}_0 &=  \frac {\vec \nabla \cdot (V' \vec B)-V'' 2 B_0 \vec B \cdot (\vec \Pi + \vec \nabla B_0) }{-2 V'' B_0^2 + V'},
\\
\dot \Pi_j &= \prt_k B_{kj}-2V' B_j.
\label{hameqs}
\eal
\eeq
As shown elsewhere \cite{Seifert:2018mmr}, 
this process reveals there are three degrees of freedom with two propagating degrees of freedom contained in two components 
of $B_j$.

Assuming boundary conditions where the $B_\mu$ and $\Pi^\mu$ 
fall off sufficiently fast at the spatial boundary, 
we integrate equation \rf{Hfin} over all space, 
and enforce the constraints (``on shell"). 
This yields the Hamiltonian,
\beq
H = \int d^3x \left(
\frac 12 \vec \Pi^2 + \frac 14 B_{ij}B^{ij} + V + 2 V' B_0^2 
\right).
\label{hamOS}
\eeq
A necessary condition for stability of the model 
is that the Hamiltonian is bounded from below \cite{Bluhm:2008yt}.
The first two terms are positive or zero, 
so we focus on the last two terms.
Note that it is the combination $V+2V'B_0^2$ that matters here, 
not the potential $V$ alone - this is due to the constraint structure 
of the model revealed through the DH process \cite{Escobar:2022lpu}.
Equivalently the same result can be obtained via the $T^{00}$ component 
of the energy-momentum tensor using a variation of the external metric, 
as shown in the next section.

To understand whether the lower boundedness condition holds for different potentials, 
we examine a number of samples, 
some of which are a key focus in this work.
Outlined in the Table \ref{pots} below are potentials and their Hamiltonian positivity properties.
We consider only smooth potentials and ignore the Lagrange multiplier potentials, 
and for simpler notation 
we define the argument of the potential to be the scalar
$X=B^\mu B_\mu$.
The classic example used frequently is the quadratic potential:
\beq
V=\frac {\la}{2} (X \pm b^2)^2,
\label{quad}
\eeq
where $b^2$ is a positive constant with the 
dimensions of mass squared.  
This potential has a minimum when $X=\mp b^2$, 
thus yielding a nonzero value for $B_\mu$ which can be spacelike, timelike,
or even lightlike.
Note that the symmetry is broken when the vacuum (quantum) state $|0>$ has the expectation value as in \rf{vac}.
Previous results have identified the quadratic potential \rf{quad} to be unstable, 
this can be seen from the Hamiltonian when the constraints are satisfied - in this case it is unbounded from below \cite{Bluhm:2008yt,Carroll:2009em}.
Other potentials have also been countenanced, 
such as a quartic potential \cite{b21}.

In a 2005 paper, 
Altschul and Kosteleck\'y showed that for an arbitrary potential for a vector field, asymptotically free theories (where at high energies the coupling constant of the nonlinear interaction term in $V$ tends to zero),
can have renormalizable and stable quantum field theories for special choices of $V$ \cite{Altschul:2005mu}.
Namely, 
these are the potentials given by, 
\beq
V=g \La^4 [ M(n,2,z) - 1 ],
\label{hyper}
\eeq
where $M(\al,\be,\ga)$ are Kummer confluent hypergeometric functions \cite{arfken},
and $z=X/\La^2$. 
%is n an integer????
Here $\La$ has dimensions of mass and represents 
the energy scale for the theory.
In this work,
we examine $n \geq 3$ which yields stable potentials with timelike 
vacuum expectation values.
We further simplify the analysis by considering $n$ to be an integer, 
for which closed form expressions for the hypergeometric functions exist.
Note that similar results with hypergeometric potentials were also found for antisymmetric tensor models \cite{abk10}.

\begin{center}
\begin{figure*}
\includegraphics[width=150mm]{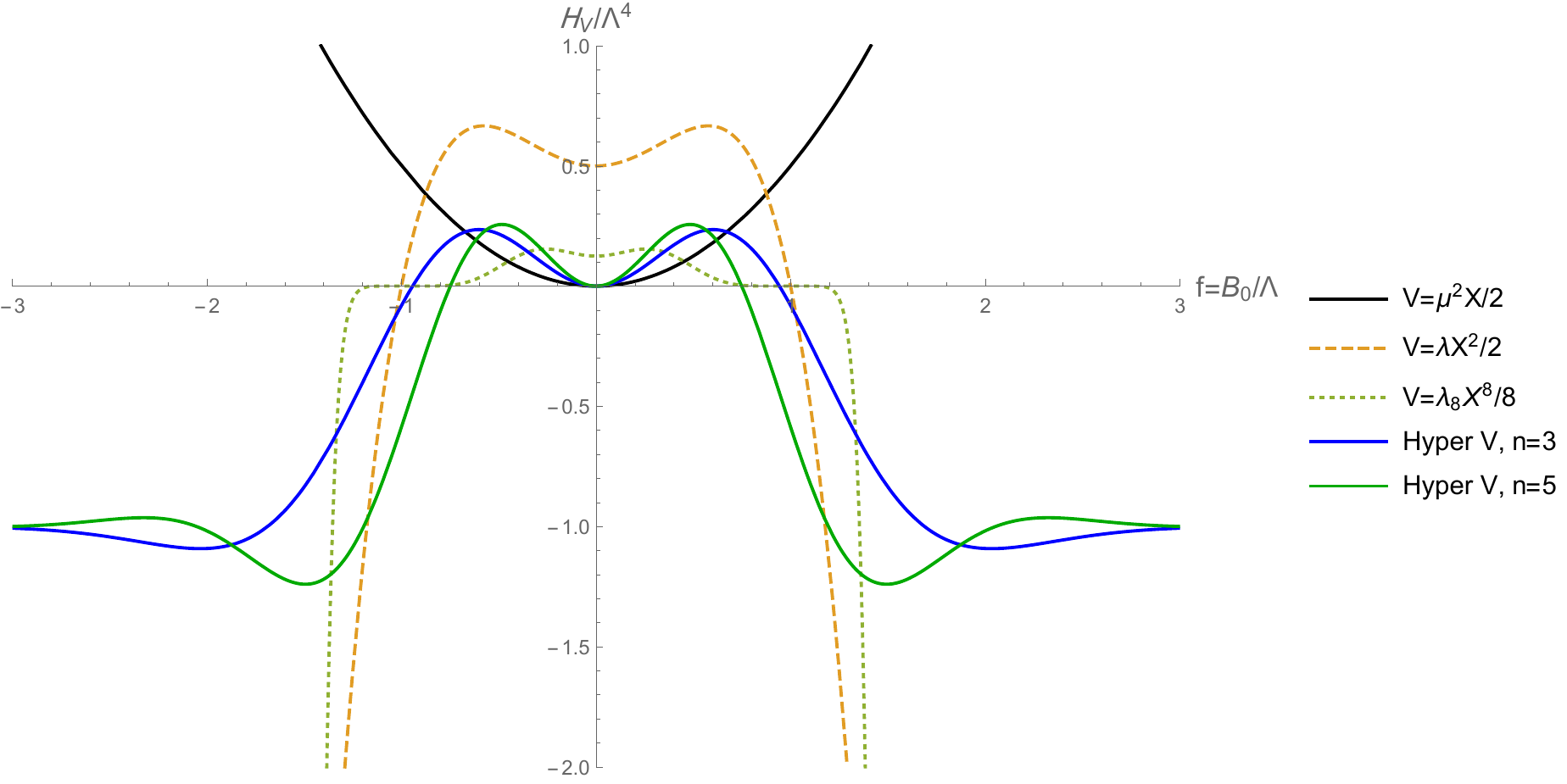}
 \caption{The potential portion of the Hamiltonian density ${\cal H}_V$ plotted for $5$ different potential functions, versus $f=B_0/\La$.  
 The black curve is the Proca potential (massive vector), 
 the dashed and dotted curves are the quadratic potential and an $X^8$ potential as polynomial samples.  Finally the blue and green curves are the hypergeometric potentials for $n=3$ and $n=5$ respectively.}
 \label{potsplot}
\end{figure*}
\end{center}

To our knowledge, 
none of these hypergeometric potentials have been considered before in a classical analysis paper, 
and not with gravity couplings.
One can confirm that for a class of hypergeometric functions, 
the Hamiltonian for such potentials is bounded from below {\it when the vector piece $||\vec B||$ or $B_0$ is fixed}.
Note that the minimum of $V$ can in fact be negative, 
which is interesting for cosmology and black holes as discussed later.
In figure \ref{potsplot} we plot the scaled dimensionless version of the potential portion 
of the Hamiltonian density from \rf{hamOS},
\beq
\frac {1}{\La^4} {\cal H}_V = \frac {V}{\La^4} + 2 \frac {V'}{\La^2} f^2,
\label{scaledHv}
\eeq
where $f=B_0/\La$.
This is done for $5$ sample potential choices including the Proca potential, 
two polynomial potentials and two hypergeometric potentials;
from which one can clearly see the behavior versus the vector field $B_0$ component.
One can also add the dependence on the spatial components $B_j$, 
which shows saddle-type behavior of the extrema
-this is also discussed elsewhere in the literature 
for other symmetry-breaking potentials \cite{Escobar:2022lpu,Kostelecky:2021xhb}. 
%However, 
%so long as $B_\mu$ is timelike, 
%the lower boundedness holds for general $B_0$ and $\vec B$.

%table of potentials
% \begin{center}
\begin{table}[!h]
    \caption{List of smooth potentials with nonzero minima for the vector field (with the exception of the Proca case), 
     with $X=B_\mu B^\mu$.  The last column indicates the boundedness of the Hamiltonian, 
     which can be checked with equation \rf{hamOS}.
     The value of $n$ is assumed to be an integer.
     These expressions also hold in curved spacetime with the minimal coupling procedure ($\et_\mn \rightarrow g_\mn$).} 
    \begin{tabular}{ l c c }
    \hline
     \hline
      & & Lower \\
      & & bound \\
      Name \,\, & Form with $X$ \,\, & on H \\
      \hline
      Proca (massive \,\, & $V=\tfrac 12 \mu^2 X$ & Yes \\ 
      \pt{ss} vector) &  &  \\
      Quadratic \,\,& $V=\frac {\la}{2} X^2$ & No \\
      Polynomial \,\,& $V=\frac {1}{n} \la_n X^n$ & No, $n$ even \\     
      Hypergeometric \,\,& $V=g \La^4 [M(n,2,X/\La^2) -1]$ & No, $n \geq 3$, \\ 
      & & Yes, fixed $||\vec B||$ \\
      \hline
      \hline
     \end{tabular}
     \label{pots}
\end{table}

To illustrate this claim on the lower boundedness consider the hypergeometric potential for the case of $n=3$, 
which has a closed form expression:
\beq
V_3=g \La^4 \left[ e^{X/\La^2}\left(1+\frac{X}{2\La^2} \right)  - 1 \right].
\label{M3pot}
\eeq
For this potential, 
the expression in \rf{scaledHv} can be written as,
\beq
\frac{1}{\La^4}{\cal H}_V
=g e^{u^2-f^2} \left(1 + u^2 \left(f^2+\frac 12 \right) 
+\frac {5}{2} f^2 -f^4 \right)-g,
\label{HVM3}
\eeq
where $f=B_0/\La$ and $u=||\vec B||/\La$.
For a vector $B_\mu$ that is timelike, 
$u^2 <f^2$, 
the limit of this expression as $f^2 \rightarrow \infty$ is $-g$ for fixed $u$, 
thus it is bounded.
For the lightlike case, 
$u^2=f^2$, 
and it can be shown that the $-f^4$ term is canceled by another term, 
leaving a positive definite expression.
For the spacelike case, 
$f^2<u^2$, 
and in this case,
it can be shown that again the term $-f^4$ is canceled by another term.

Despite this analysis, there remains a subtle aspect overlooked in a previous version of this manuscript.\footnote{We thank Robertus Potting for pointing this out.}
This occurs along hyperbolas defined by constant $t=u^2-f^2$.
In this case the hamiltonian is proportional to $e^t [1+f^2 (3+t)+\tfrac 12 t]-1$.
From this expression for fixed $t<-3$, 
the Hamiltonian can become arbitrarily negative for large $f$.  
Thus, the Hamiltonian for the hypergeometric function is {\it not bounded from below in the generic $B_\mu$ case}.
Nonetheless, for subsets of the $B_\mu $ space, like constant $u$, it is bounded.
Our focus in this work is on the latter.

A study of the local extrema of this function \rf{HVM3} with respect to $(f,u)$, 
shows that they exist at the points $f=\pm \sqrt{9\pm \sqrt{57}}/2, u=0$ and $f=0,u=0$. 
The value of ${\cal H}_V/(g\La^4)$ at these extrema is $0.235$ when $f=\pm \sqrt{9 - \sqrt{57}}/2$,
$-1.092$ when
$f=\pm \sqrt{9 + \sqrt{57}}/2$, 
and $0$ when $f=0=u$.
These extrema, which are all timelike values of $B_\mu$ turn out to be saddle points except for the solution $B_\mu =0$, 
which is a local minimum.

This analysis can be carried out for the $n>3$ potentials as well.
We have checked for the same properties for the $n=4$, $n=5$ hypergeometric potentials, 
but a thorough study along these lines of the generic hypergeometric potentials remains open.
In particular, 
conclusions for any $n$ may be drawn using the asymptotic properties of the hypergeometric functions \cite{Altschul:2005mu}.

Another aspect we mention here is that the Hamiltonian \rf{Hfin}
can be used to verify that a static solution with only $B_0 \neq 0$
will remain static under time evolution, 
and that no $B_j$ arises.
In this static limit, 
the Hamilton's equations from above reduce to
\beq
\bal
\dot{B}_0 &=0, \\
\dot{\Pi}^0 &=\prt_j \Pi^j + 2 V' B_0=0 \\ 
\Pi^j &= -\prt_j B_0, \\
\dot{\Pi}^j &=0.\\
\label{staticEvol}
\eal
\eeq
Using the time evolution equation for each of the four equations above $\dot{F}=\{F,H\}$,
it is straightforward to show that these static equations will hold for all $t$.

\subsection{Flat spacetime solutions}
\label{solutionsflat}

\subsubsection{Linearized solutions}
\label{linearized solutions}

Returning to \rf{flatBB} or the Hamilton equation \rf{SC}
and assuming a timelike vector field aligned in the chosen coordinates 
as $B_\mu = (B_0, \vec 0)$,
the time component of the field equations is
\beq
\nabla^2 B_0 = 2 V' B_0 
\label{flatBB0}
\eeq
and the potential dependency is 
$V=V(-B_0^2)$.
Denote the (vacuum) solution $b_\mu = (\pm b,0,0,0)$, 
where $b \geq 0$,
as that which minimizes $V$ and hence satisfies $V'=0$.
When expanded around this solution $B_\mu = b_\mu + \be_\mu$, 
assuming $\prt_\nu b_\mu=0$ and seeking an equation linear in $\be_0$,
equation \rf{flatBB0} becomes
\beq
\nabla^2 \be_0 = - 4 {\bar V}'' b^2 \be_0,
\label{flatBBsph}
\eeq
where the bar indicates evaluation with $b_\mu$.
Since ${\bar V}''>0$ the equation is just
the standard homogeneous Helmholtz equation. 
The (spherically symmetric) solutions are 
\beq
\be_0 = \frac {1}{r} \left[c_1 \cos (k r) + c_2 \sin (kr) \right],
\label{linsolns}
\eeq
where $k=2b\sqrt{ {\bar V}''}$.
Note that the second term is a spherical Bessel function $j_0 (kr)$;
furthermore the general solution can include angular dependence we have ignored here( see Refs.\ \cite{Jackson} or \cite{arfken}).

The asymptotic behavior of \rf{linsolns} is such that $\be_0 \rightarrow 0$
as $r \rightarrow \infty$, 
so that the field $B_0$ relaxes to its vacuum value at spatial infinity.
We will use this property and \rf{linsolns} below.
 Finally we note that the solution \rf{linsolns} produces a radial electric field $E_r=F_{r0}=\prt_r B_0$,
given by
\beq
\bal
E_r &= -\frac {1}{r^2} 
\left[c_1 \cos (k r) + c_2 \sin (kr) \right] \\
& 
\pt{sp}
+\frac {k}{r} 
\left[ - c_1 \sin (kr) + c_2 \cos (kr) \right].
\label{E}
\eal
\eeq
Note that wave-like $1/r$ terms show up in this modified electrostatic result.
Our results here are consistent with the analysis of Ref.\ \cite{bk08}.
In that paper, 
it was shown that $\be_0$ is nonpropagating and represents the solution to an initial value constraint equation rather than a dynamical equation.
However, 
we do not include a time-dependent vector component $B_j$, 
so the solutions differ slightly.

Given the result \rf{E}, 
one could pursue a program of phenomenology searching for effects of the extra terms in static situations.
Indeed, 
it was shown long ago
that massive photon modifications drastically change electrostatics
and have been used to tightly constrain the photon mass
(e.g., $m_{\ga}< 3\times10^{-10} \, cm^{-1}$ \cite{Williams71}).
On dimensional grounds, 
we might expect similar constraints on $k$.
Also, 
it was shown that similar modifications can arise in short-range gravity tests, coming
from generic Lorentz-breaking terms, 
with higher mass dimension, 
in the action of an EFT framework \cite{Bailey_2023}.
To pursue this work for the bumblebee model interpreted as a modified electrodynamics model, 
we leave for the future.

\subsubsection{Spherical symmetry case, nonlinear solutions}
\label{spherical nonlinear}
% \subsection{Uniqueness, 
% trajectory stability and symmetry breaking attractors}
% %sounds cool anyway...

Assuming spherical symmetry ($B_0 = B_0 (r)$) we have for \rf{flatBB0},
\beq
\frac{1}{r^2} \frac {d}{dr} \left(r^2 \frac {d B_0}{dr} \right) = 2 V' B_0. 
\label{sph}
\eeq
We define a dimensionless coordinate $x=\La r$
and divide this equation by $\La$; using $f=B_0/\La$.
The dimensionless form of the equation is then
\begin{equation}
    \frac{d^2f}{dx^2}+\frac{2}{x}\frac{df}{dx}=2fV'(-f^2).
    \label{Dimensionless Spherical Symmetry Form}
\end{equation}
This equation is an ordinary differential equation (ODE) that is second order, nonlinear, and nonautonomous (has explicit $x$ dependence).
For possible potential choices we refer to Table \ref{pots}.
We begin studying the solution to this equation by noting several variable substitutions that can be useful for studying different aspects of the solutions.
Calling $2fV'(-f^2) = h(f)$, 
we consider two alternate forms:
\begin{equation}
      x=\frac{1}{u} \:\:\rightarrow \:\: u^{4}\frac{d^2f}{d^2u}=h(f(u)),
      \label{usub}
\end{equation}
\begin{equation}
     x=e^{\alpha t}\:\:\rightarrow \:\: \frac{d^2f}{dt^2}+\alpha\frac{df}{dt}=\alpha^2e^{2\alpha t}h(f(t)),
     \label{tsubs}
\end{equation}
where the domain $0 \leq x \leq \infty$ maps $1-1$ to the corresponding range for $u$ and $t$;
namely $0 \leq u \leq \infty$, $-\infty \leq t \leq \infty$, respectively.

To get an idea of the challenge of finding analytical solutions consider the quadratic potential case, 
using \rf{tsubs}:
\beq
\ddot{f} + \al \dot{f} = 2 \al^2 \la e^{2\al t} \la (1-f^2)f. 
\label{quadt}
\eeq
An exhaustive source for analytical solutions to ODEs, including nonlinear ones is given in Ref.\ \cite{polyanin}. Amongst the equations therein, 
the closest match to \rf{quadt} is the following equation:
\beq
\ddot{y} -\al \dot{y} = e^{2\al t} f(y),
\label{2.9.2.17}
\eeq
where we have changed dependent variables from $f \rightarrow y$ to distinguish the equations and solutions.
Note this equation differs only by one critical sign from \rf{quadt}.
This equation has an analytical solution:
\\
\beq
\int_{}^{} \frac{dy}{\sqrt{C_1 + 2  \int_{}^{} f(y) \,dy }} = C_2 \pm \frac{1}{\al}e^{\al x}.
\label{2.9.2.34 Solution}
\eeq
Unfortunately, 
attempts to find a similar exact solution for the quadratic potential case \rf{quadt} fail.

\subsubsection{Stability argument and numerical solutions}
\label{analyticargument}
%lets add M4 plot also

Using the substitution found in \rf{usub} can lead to a generalized statement about the behavior of the function for a given potential.
The differential equation becomes,
\beq
u^4\frac{d^2f}{du^2}=2fV'(-f^2).
\label{Stability Argument}
\eeq
Since $u^4=\frac{1}{x^4}$ is always positive, 
the sign on the left-hand side corresponds to the sign of $\frac{d^2f}{du^2}$, and the initial concavity of the solution. 
The behavior for three different potentials 
are shown in the equations,
\beq 
\label{Quadratic Stability Argument} u^4\frac{d^2f}{du^2}=2\lambda f(1-f^2) \rightarrow \left\{ 
\begin{array}{ll}
      f_{uu}>0 &:\; f<-1 \\
      f_{uu}<0 &:\; -1<f<0 \\
      f_{uu}>0 &:\; 0<f<1 \\
      f_{uu}<0 &:\; f>1 \\
\end{array}
\right\},
\eeq

\beq 
\label{Hypergeometric 3 Stability Argument} 
u^4\frac{d^2f}{du^2}=ge^{-f^2} (3-f^2)f \rightarrow \left\{ 
\begin{array}{ll}
      f_{uu}>0 &: f<-\sqrt{3} \\
      f_{uu}<0 &: -\sqrt{3}<f<0 \\
      f_{uu}>0 &: 0<f<\sqrt{3} \\
      f_{uu}<0 &:f>\sqrt{3} \\
\end{array}
\right\},
\eeq

\beq 
\bal
\label{Hypergeometric 4 Stability Argument} u^4\frac{d^2f}{du^2}&=\frac{1}{6}g e^{-f^2} f \left(f^4-8 f^2+12\right)\\
&\rightarrow \left\{ 
\begin{array}{ll}
      f_{uu}<0 &:\; f<-\sqrt{6} \\
      f_{uu}>0 &:\; -\sqrt{6}<f<-\sqrt{2} \\
      f_{uu}<0 &:\; -\sqrt{2}<f<0 \\
      f_{uu}>0 &:\; 0<f<\sqrt{2} \\
      f_{uu}<0 &:\; \sqrt{2}<f<\sqrt{6} \\
      f_{uu}>0 &:\; f>\sqrt{6} \\
\end{array}
\right\}.
\eal
\eeq

Equation \rf{Quadratic Stability Argument} shows the stability argument for the quadratic potential,
Eq.\ \rf{Hypergeometric 3 Stability Argument} applies the same reasoning to the hypergeometric potential $V=g \La^4 [M(3,2,z) -1]$, 
and Eq.\ \rf{Hypergeometric 4 Stability Argument} applies to  $V=g \La^4 [M(4,2,z) -1]$, 
where $z=X/\La^2$.
Here,
$f_{uu}$ denotes the second derivative of $f$ with respect to $u$, $\frac{d^2f}{du^2}$.
%Note that we can apply an argument about the trend of $f$ as $u$ tends toward smaller values because of the symmetry of the equations in \rf{Stability Argument} under the exchange $u\rightarrow -u$.

\begin{figure*}
\includegraphics[width=120mm]{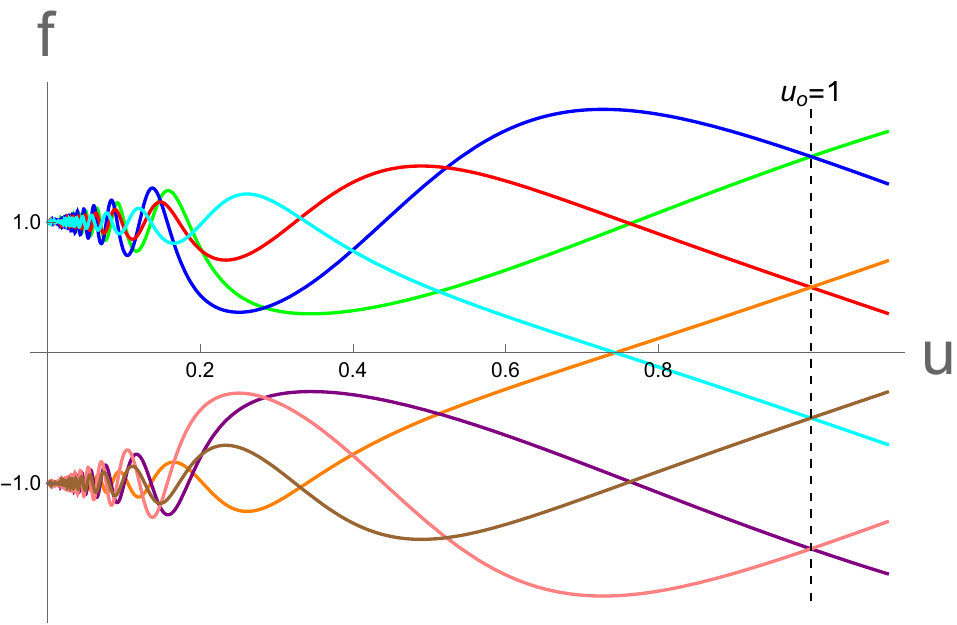}
 \caption{Plot of a series of numerical solutions for the quadratic potential $V=\frac {\la}{2} X^2$ with respect to $u=\frac{1}{x}$ with each solution showing dampening oscillations at $\pm1$ as $u$ approaches 1, or $x$ approaches infinity.  We set $\la=1$ for this plot.
The colors represent distinct initial values for $f$ and $df/du$ at the point $u_0=1$.
 }
 \label{Quadratic Stability pdf}
\end{figure*}

In the quadratic potential case, 
whenever $f \neq 1$ 
the concavity of $f(u)$ changes to move $f$ towards $f=1$ or $f=-1$.
Thus $f_{uu}<0$ when $f>1$
and the $f_{uu}>0$ when $f<1$. 
This behavior occurs regardless of the initial position or initial slope $f_u$, and the solution asymptotes to one of the two points of stability $f=\pm1$. 
The plot in Fig.\ \ref{Quadratic Stability pdf} shows how this argument applies regardless of the initial conditions. 

\begin{figure*}
\includegraphics[width=120mm]{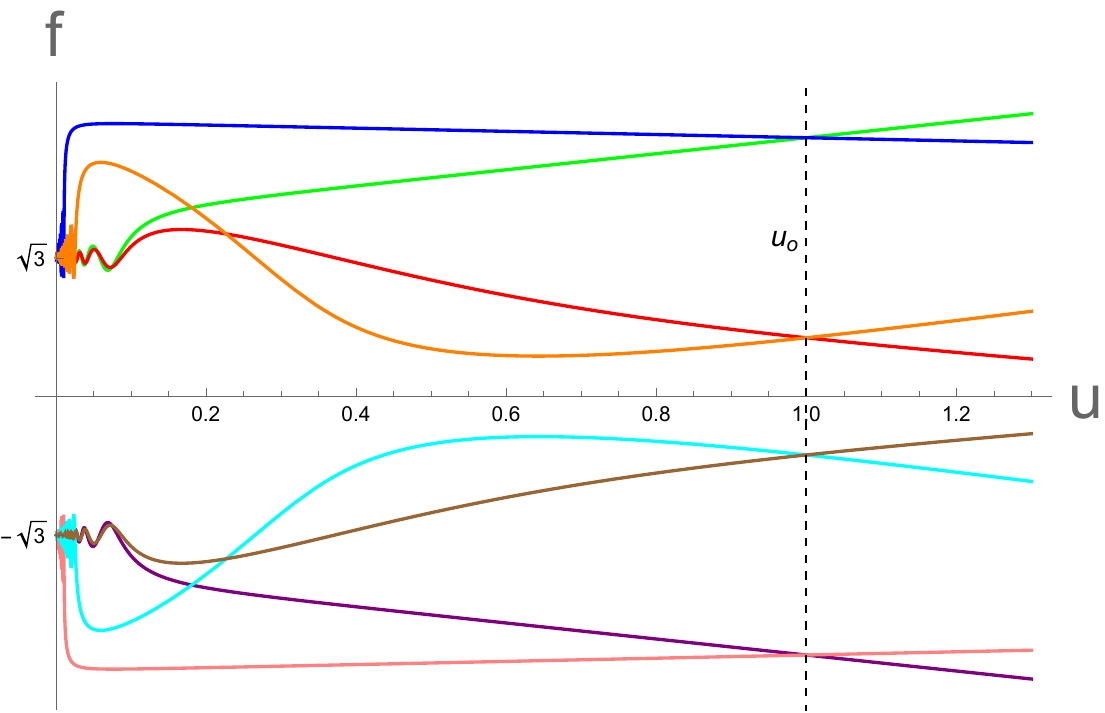}
 \caption{Plot of a series of numerical solutions for the hypergeometric potential $n=3$, with each solution showing dampening oscillations around $\pm\sqrt{3}$.  We set $g=1$ for this plot.  The colors represent distinct initial values for $f$ and $df/du$ at the point $u_0=1$.}
 \label{n3 Hyper Stability pdf}
\end{figure*}

\begin{figure*}
\includegraphics[width=120mm]{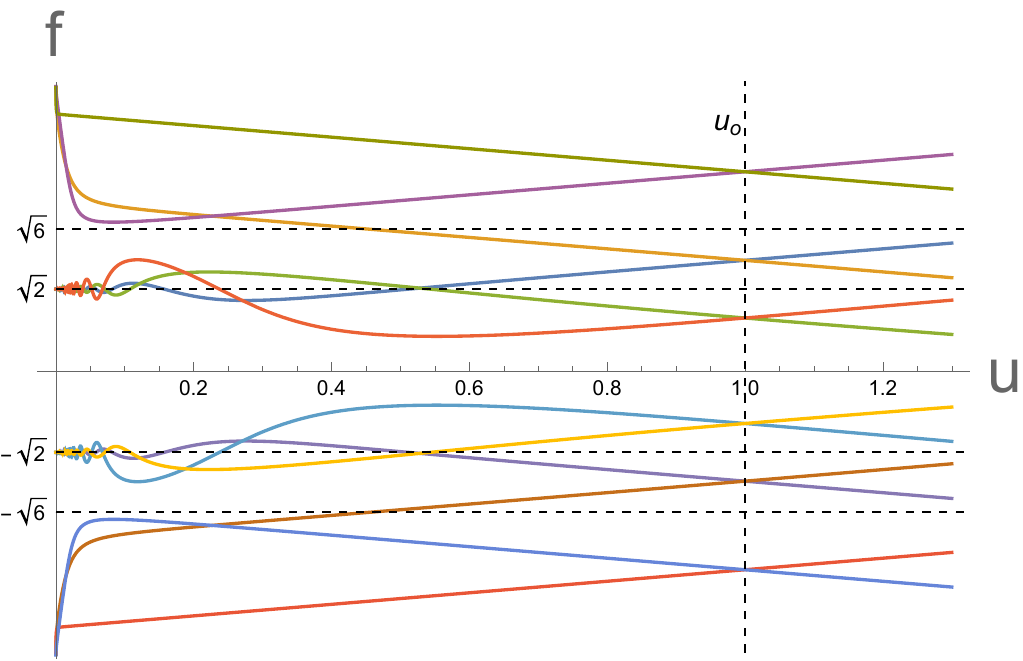}
 \caption{Plot of a series of numerical solutions for the hypergeometric potential when $n=4$, with some solutions showing dampening oscillations around $\pm\sqrt{2}$, but solutions that move a distance of $\sqrt{6}$ from the origin escape completely as $u$ approaches $0$, or as $x$ approaches infinity.  We set $g=1$ for this plot.  The colors represent distinct initial values for $f$ and $df/du$ at the point $u_0=1$.}
 \label{n4 Hyper Stability pdf}
\end{figure*}

Identical behavior occurs for the hypergeometric potential with the $n=3$ case but the stable points are $\pm \sqrt{3}$.  
This is demonstrated in Fig.\ \ref{n3 Hyper Stability pdf}.
However, 
with the hypergeometric $n=4$ case, 
when $f< -\sqrt{6}$ or $f>\sqrt{6}$, 
the concavity of the solution moves it only further away from the u-axis, 
and thus it does not stabilize, 
which is illustrated in Fig.\ \ref{n4 Hyper Stability pdf}. Overall then, 
we find that the first two potentials, 
quadratic and hypergeometric ($n=3$) are asymptotically stable, 
while $n=4$ is not.

The argument used here applies when following decreasing values of $u$, 
which represents increasing values of $x$. 
What happens on the ``other end", 
when one approaches the origin?
When looking at increasing values of $u$ or $x$ near the origin, 
we consider the following limit:
\begin{equation}
    \lim_{u\rightarrow\infty}\frac{d^2f}{du^2}=\lim_{u\rightarrow\infty}\frac{h(f(u))}{u^4},
    \label{u limit}
\end{equation}
where $h$ is the right-hand side of Eqns.\ (\ref{Quadratic Stability Argument}-\ref{Hypergeometric 4 Stability Argument}).
In the case of the hypergeometric potentials, 
the function $h(f)$ is bounded, 
due to the $e^{-f^2}$, 
such that $|h(f)| \le \ep$ for some positive constant $\ep$.
We can then conclude that $f_{uu}$ approaches $0$. This means that $f_u$ approaches some constant, 
suggesting 
that the solution diverges as $u$ increases, or as $x$ approaches $0$.
This is similar to the electrostatic case with $V'=0=V$ which has the exact solution $f=au+b$ for some constants $a$ and $b$.

\section{Coupled gravity-vector solutions, Part I}
\label{gravity-vector, part I}

This section contains a broad study of the field equations when gravity is present, 
confining attention to the effects of the potential $V$
as in equation \rf{xi0}.
First we display the field equations for gravity and the bumblebee field.
Variation of the action in equation \rf{bbmodel}, 
with respect to the metric $g_\mn$ and the field $B_\mu$, 
yields
\beq 
\bal
G_\mn &= \ka \Big[ B^\al_{\pt{\al}\mu} B_{\al\nu} - \frac {1}{4} g_\mn B^\ab B_\ab \\
&\pt{space} 
-V g_\mn + 2 V' B_\mu B_\nu \Big], \\
\nabla^\mu B_\mn &= 2 V' B_\nu,
\label{FEgravBB}
\eal
\eeq
where $G_\mn$ is the Einstein tensor.
These equations can be confirmed from Ref.\ \cite{k04} and others. 
Note that the antisymmetry of the field strength tensor $B_\mn$ implies the current conservation identity
\beq
\nabla^\nu (2 V' B_\nu)=0,
\label{current}
\eeq
which must hold for any solutions.
Since the theory is fully dynamical, $\nabla^\mu T_\mn =0$ holds when the vector field equations are satisfied, 
where $T_\mn$ is the bracketed expression in \rf{FEgravBB}.
That this holds in a spontaneous spacetime symmetry-breaking model is expected \cite{k04,bluhm15}.

We can make some general remarks on the weak and null energy conditions at this stage \cite{Carroll:2004st}.
The first part of the stress-energy tensor (the first two terms on the right-hand side of the $G_\mn$ part of equations \rf{FEgravBB}
is the standard electromagnetic piece, 
present in Maxwell theory.
This can be shown to satisfy both the weak and null energy conditions, 
a widely known result.
Thus, 
consider now the potential part
$T_V^\mn = -Vg^\mn + 2V' B^\mu B^\nu$.
Contraction on both indices with a unit timelike vector $u_\mu$ yields
\beq
T_V^\mn u_\mu u_\nu  = V+2V' (B_\mu u^\mu )^2.
\label{weak}
\eeq
This is the arbitrary spacetime generalization of the potential portion of the Hamiltonian density \rf{scaledHv}.
So we know that for hypergeometric potentials this is bounded from below in flat spacetime, 
when $||\vec B||$ is fixed, but it is unbounded in the general case.
These statements are valid in a local frame in the neighborhood of a point but can be demonstrated for all points.
Positivity of \rf{weak} fails for many other potential choices, with the exception of the massive potential.
So it appears that in many cases, 
the weak energy condition is violated.

For the null energy condition, 
using a null vector $k_\mu$, 
we find
\beq
T_V^\mn k_\mu k_\nu = 2 V' (B^\mu k_\mu)^2.
\label{null}
\eeq
It is now unclear if the right-hand side is positive or zero, 
since the dependence is on $V'$. 
Since $V$ is chosen to have a minimum for some value of $X$,
$V'$ can have positive and negative values away from that minimum.
Thus, 
it appears there could be a violation of the null energy condition.
Energy conditions in other vector models of spacetime-symmetry breaking have been studied in Ref.\ \cite{Garfinkle:2011iw}.

In this section, 
the plan of study is as follows.
First, 
we discuss an exact solution for the case
of constant vector field at the minimum of the potential in cosmology.
Next we study the full equations for the spherically symmetric spacetime case, 
assuming the vector field lies away from the minimum.
We then explore how the model compares to other beyond-GR models with extra fields to see if Bekenstein's horizon argument holds (this will allow us to see if black hole solutions are possible or reduce to trivial cases).
We then discuss some exact solutions that can be obtained under certain conditions on the fields.
The section that follows this one explores generic solutions using series expansions, numerical code, 
and arguments from curvature invariants.
Orbits in spacetimes described by these solutions are also explored.

\subsection{Exact anti-de Sitter cosmological solutions}

It was previously shown that the bumblebee model can yield de Sitter or Anti de Sitter (AdS) solutions depending on the potential minimum value \cite{Jesus:2019nwi}.
To show that the model produces AdS \cite{Bousso:2002fq} solutions for the hypergeometric potentials, 
%add antidesitter reference
we make the following assumptions on the vector field and the metric in spherical coordinates $\{t,r,\th,\ph\}$:
\beq
\bal
B_\mu &= (b,0,0,0), \\
g_\mn &= {\rm diag} (-1,a^2,a^2 r^2,a^2 r^2 \sin^2 \th). \\
\label{FLRWfe}
\eal
\eeq
We are thus assuming a spatially flat FLRW cosmology
ansatz with an isotropic vector field.
It is straightforward to show with \rf{FLRWfe}, 
that if the potential lies at its minimum $V'=0$, 
the field equations reduce to
\beq
G_\mn = -\ka \bar{V} g_\mn,
\label{FLRW2}
\eeq
where $\bar{V}$ is the potential evaluated at its minimum.  
This result then implies, 
via $G_\mn = -\La_c g_\mn + ...$,
an effective cosmological constant,
\beq
\La_c = \ka \bar{V}.
\label{Lambda}
\eeq
Here we use the subscript $\La_c$ to indicate the cosmological constant rather than the energy scale $\La$ used earlier.

Glancing at the potentials in Table \ref{pots}, 
the only choices with nonzero minima for $V$
are the hypergeometric potentials.
For example, 
for $V=g\La^4[ M(3,2,z)-1]$, 
the two minima for $B_0$ occur when 
\beq
\bal
X &= -3 \La^2, \\
\bar{V} &=-g \La^4 \left(1+ \frac {1}{2 e^3} \right),
\label{min3}
\eal
\eeq
thus indicating a naturally occurring negative cosmological constant for the hypergeometric potentials.
By natural we mean that the minimum value arises from the inherent nature of the potential itself, 
not by adding an {\it ad-hoc} constant to the action, 
like the usual $\La_c$.
This implies that anti-de Sitter spacetime \cite{Sokolowski:2016tar} can arise from the hypergeometric potentials in the bumblebee model.
Indeed, 
the Friedmann equations arising from equation \rf{FLRW2} yield
\beq
\bal
\frac {\ddot{a}}{a} = -\frac {1}{3} \ka |\bar{V}|, 
\label{friedman}
\eal
\eeq
with the general solution
\beq
a(t) = a_1 \sin \frac {t}{\ta} + a_2 \cos \frac{t}{\ta},
\label{cosmo}
\eeq
where $\ta = \sqrt{3/\ka \bar{V}}$.

This type of solution \rf{cosmo} is an oscillating universe.
Let us assume that the time scale is of the order of the age of the universe $\ta \approx 10 \, Gyr \approx 3\times 10^{17} s$.
Using the form of the (hypergeometric) potential above, 
$\ta = 5 \times 10^{-43} s/(k \sqrt{g})$, 
where $k=\ka \La^2 \sim \La^2/M_{Pl}^2$ is the ratio of the mass scale of the bumblebee model to the Planck mass squared. 
To match the observed time scale of the universe we see
that
\beq
\sqrt{g}k \approx 10^{-60},
\label{finetune}
\eeq
which is incredible fine-tuning and represents nothing more than the usual cosmological constant problem \cite{RevModPhys.61.1,Martin:2012bt}.

When we turn to spherical coordinates and static spacetime, 
we will still see the evidence of the same kind of solution.
This is because it is known that the AdS solution \rf{cosmo} can be mapped via a coordinate transformation to a static and spherically solution with a term of the form $g_{tt} \supset r^2/l^2$ for some length scale $l$ \cite{Bousso:2002fq}.
To avoid this, 
in Sec.\ \ref{gravity-vector part II} 
we will subtract the minimum off of the potential for the hypergeometric potentials.
We do not make this assumption in the remainder of this section, however.
%more cites here?

\subsection{Full equations and general features} 
\label{full equations}

The goal is to seek solutions for the model \rf{FEgravBB}
in the spherically symmetric static case.
We assume the metric and bumblebee field take the form
\beq
\bal
ds^2 &= A \, dt^2 +B \, dr^2 +r^2 d\Om^2, \\
B_\mu &= (B_0,0,0,0),
\label{ansatz2}
\eal
\eeq 
where $A$, $B$, and $B_0$ are functions of the coordinate $r$ only.
Note that one could also explore solutions with $B_r \neq 0$ and still be consistent with spherical symmetry,
as done with the nonminimal coupling $\xi_2$ in other works.
We leave this for future work to explore this option with off potential minimum solutions.

Upon imposing \rf{ansatz2} on the gravity fields equations \rf{FEgravBB}, 
we obtain the following nonzero components:
\beq
\bal
G_{tt} &= -\frac {A B'}{B^2 r} - \frac{A}{r^2}\left(1-\frac{1}{B} \right) \\
&= \ka \left( \frac{B_0^{\prime 2}}{2 B} -V A + 2 V' B_0^2
\right), \\
G_{rr} &= \frac{A'}{A r} + \frac {1-B}{r^2} = \ka \left(
\frac{B_0^{\prime 2}}{2A} - V B
\right),\\
G_{\th\th} &= \frac {r}{2B} \left(\frac{A'}{A} - \frac {B'}{B} \right)- \frac {r^2 A'}{4 A B}
\left(\frac{A'}{A} + \frac {B'}{B} \right) + \frac{r^2 A''}{2 B A}
\\
&=-\ka \left( \frac {r^2 B_0^{\prime 2}}{2 A B} + r^2 V    \right),\\
G_{\ph\ph} &= \sin^2 \th G_{\th\th}.
\label{sphgrav}
\eal
\eeq
For the vector field equations we find the only nonzero component is 
\beq
B_0^{\prime \prime} + \frac{2}{r}B_0'-\frac {B_0'}{2} \left(\frac{A'}{A} + \frac {B'}{B} \right) =2V'B B_0. 
\label{sphvec}
\eeq
These equations are coupled nonlinear ODEs.
Not all the equations are independent.
The $G_{\ph\ph}$ equation is redundant, 
and among the three equations for $G_{tt}$, $G_{rr}$, 
and $G_{\th\th}$, 
there are only two independent ones.
This is due to the traced Bianchi identities $\nabla_\mu G^\mn =0$.
Notice that $G_{\th\th}$ contains 
the second derivative $A''$; 
to make the analysis simpler, 
we can opt for the two first order equations $G_{tt}$ and $G_{rr}$
instead.
Of course, 
the second order equation for $B_0$
cannot be avoided, 
as we need three equations for three unknowns.

Combining $G_{tt}$ and $G_{rr}$ we obtain,
\beq
\left(\frac{A'}{A} + \frac {B'}{B} \right)=
-2\ka V' B \frac {B_0^2}{A}r,
\label{identity}
\eeq
which can then be re-inserted back into the vector equation \rf{sphvec}
to simplify the expression.
We also solve the $G_{tt}$ and $G_{rr}$ for $A'$ and $B'$.
All together we can write the three equations we study in the (mostly) first order form,
\beq
\bal
B_0^{\prime \prime} + \frac{2}{r}B_0'
&= 2V'B B_0 -\ka V' B \frac{B_0^2}{A} B_0'r,\\
B' &= \frac {B}{r} (1-B)  \\
&\pt{= }
+ \ka \left( V B^2 -\frac{B B_0^{\prime 2}}{2 A} - 2 V' \frac{B_0^2}{A} B^2 
\right)r, \\
A' &= \frac {A}{r} (B-1) + \ka 
\left(
\frac{B_0^{\prime 2}}{2} - V A B
\right)r.
\label{firstorder}
\eal
\eeq
The Reissner-Nordstr\"om (RN) solution is obtained for the case when $V=V'=0$:
\beq
\bal
B_0 &= c_1 + \frac{c_2}{r}, \\
A &= -1 + \frac{c_3}{r} - \frac{\ka c_2^2}{2 r^2},\\
B &= -\frac{1}{A},
\label{RNsoln}
\eal
\eeq
where the $c_n$'s are fixed by the values of $B_0$, $B_0'$, and
$A$ at some position $r_0$.
We will show below that exact solutions related to equation \rf{RNsoln} are possible with a nonzero $V$.

\subsection{Bekenstein horizon condition}
\label{bekenstein}

It is known that for Brans-Dicke theory and the massive vector field case, 
as well as others, 
there are no static solutions with horizons except for the trivial case when the scalar or vector field is constant \cite{Sotiriou:2011dz,Obukhov:1999ed,Bekenstein:1971hc,Bekenstein:1972}.
Indeed, 
this is part of the ``no-hair" conjecture, 
which says that stationary black holes are characterized only by their mass, 
charge, and angular momentum.
Exceptions to this theorem, 
while exotic and not corresponding to any Standard Model fields, abound \cite{Herdeiro:2015waa,Mavromatos:1995fc}.

In practice, 
one uses a suitable integration of the field equations for a scalar or vector field 
to establish if a solution with a horizon can depend on quantities other than 
the suitably defined mass, angular momentum, and electric charge.
In the case of the bumblebee model, 
we take the contraction of equation \rf{FEgravBB} with $B^\nu$ and examine the integral 
over a four-dimensional volume of spacetime
${\cal V}$:
\beq
\int_{\cal V} d^4 x \sqrt{-g} \left( B_\nu \nabla_\mu B^{\mu\nu} - 2 V' B_\nu B^\nu  \right)=0.
\label{int1}
\eeq
Integrating by parts leaves a surface term:
\beq
\bal
& \int_{\prt \cal V} d^3 \Si_\nu
\sqrt{-g} B_\mu B^\mn 
\\
&- \int d^4x \sqrt{-g} 
\left( \frac 12  B_\mn B^\mn + 2 V' B_\nu B^\nu  \right)
=0.
\label{int2}
\eal
\eeq
At this stage one specializes the four dimensional volume ${\cal V}$ to be that bounded by 
two Cauchy (spacelike) hypersurfaces at different times, 
a portion of the black hole horizon, 
and a timelike three-surface at spatial infinity.

The surface terms vanish due to the following:
the integrals over the Cauchy hypersurfaces cancel each other when we assume 
a static spacetime;
the asymptotic flatness assumption
ensures that $B_\mn \rightarrow 0$ at spatial infinity.
The vanishing of the integral over the horizon is more subtle.
The integral involves the scalar
$d^3\Si_\mu \ze^\mu$, 
where $\ze^\mu = F^\mn B_\nu$.
Bekenstein has shown that this integral will vanish on the horizon provided that $\ze^\mu$ is bounded on the horizon.
We do not show the details here; the reader can refer to the original papers mentioned above.

It suffices for our work to argue that $\ze_\mu \ze^\mu$, 
which is spacelike on account of $F^\mn B_\mu B_\nu=0$, 
is bounded on the horizon.
Now, there should be no singularities of (physical) tensor quantities on the horizon, 
so $T_\mn$ must be regular.
For example, 
the trace $T_\al^{\pt{\al}\al}=-4V+2V'X$, 
where $X=g^\mn B_\mu B_\nu=B_0^2/A$
with the spherical ansatz in \rf{ansatz2}.
For the generic potential $V$, 
if $X$ is to be regular on the horizon then we must have $B_0=0$ on the horizon.
This then implies that $\ze_\mu \ze^\mu$ also vanishes on the horizon since it is proportional to $B_0^2$.
Therefore $\ze_\mu \ze^\mu$ is bounded, 
and the horizon surface integral vanishes.
This argument is similar to that used for the Einstein-Proca model in Ref.\ \cite{Obukhov:1999ed}.

This leaves the remaining volume integral which must be zero.
Assuming the form
\rf{ansatz2} yields,
integrating out the time interval
$\de t$ 
since the fields are static, 
we obtain
\beq
\int d^3x \sqrt{g} \frac {1}{|A|} \left(\frac {B_0^{\prime 2}}{B} + 2V'B_0^2
\right)=0, \,\,\, {\rm (if \, a \,  horizon \, exists) }
\label{int3}
\eeq
where we assume that $A<0$ and $B>0$.

The standard argument involves establishing in equation \rf{int3}
that both terms in the integrand are positive definite and therefore must vanish since the integral over a spatial volume vanishes.
For example, 
for the Proca vector field case
with $V=\mu^2 B_\mu B^\mu/2$, 
we get that $V'=\mu^2/2 \geq 0$
and so both terms in \rf{int3} must 
independently vanish.
This then implies that $\prt_r B_0=0$
and thus the only way have a horizon in this model is the trivial case for $B_0$ that reduces to the Schwarzschild solution.

However for the other choices of 
potential in Table \ref{pots}, 
$V'$ is indefinite.
Consider the case of the hypergeometric potential with $n=3$.
In this case we get
\beq
V' = \frac {g}{2} \La^4 e^{-f^2/|A|} \left(3- \frac {f^2}{|A|}\right).
\label{hypv'}
\eeq
Here we can see that this term can be positive or negative depending on the value of $f=B_0/\La$, 
so it is not immediately obvious that this result \rf{int3} implies any constraints on the bumblebee model in this limit.
It appears that this model (and others) evade the Bekenstein ``no-hair" result.
Note, 
however, 
that solutions without horizons
also 
escape this argument, 
so it is possible 
to have naked singularity solutions which we find in Section \ref{gravity-vector part II}.

\subsection{Some specialized exact solutions}
\label{special solutions}

Some special solutions can be countenanced for the equations in
\rf{firstorder}.
They stem from assuming the condition $(AB)'=0$. This condition implies $AB=\al$ where $\al$ is a constant. By multiplying equation \rf{identity} by $AB$, we see that
\beq
-2\ka V'B^2B_0^2r=0.
\eeq
Because this equation must be true for all $r$, we must have either $B_0=0$ or $V'=0$ (we are not considering a case where $B=0$ for all $r$).

In the former case, the fact that $B_0=0$ for all $r$ means that $V$ is constant as $V$ is a function of $B_0^2/A$
(denoted $V_0 = V(0)$ to distinguish from $\bar V$). 
We can use $B=\al/A$ to rewrite the first order equations in \rf{firstorder} into just one equation with $A$, which is
\beq
A'=\frac{1}{r}(\al-A)-\ka\al V_0 r,
\eeq
with solution,
\beq
A=\al+\frac{c_1}{r}-\frac{\ka\al V_0 r^2}{3},
\label{SAdS}
\eeq
where $c_1$ is a constant.
With $\al=-1$, 
this solution is known as the Schwarzschild-de Sitter or Schwarzschild-AdS solution \cite{PhysRevLett.100.171101}, 
depending on the sign of $V_0$.

In the latter case, where $V'=0$, we can rewrite the first order equations in \rf{firstorder} to
\beq
\bal
B_0''+\frac{2}{r}B_0'&=0\\
A'&=\frac{1}{r}(\al-A)+\ka\left(\frac{B_0'^2}{2}-\al V\right) r,
\label{Vp0case}
\eal
\eeq
where the equation for $B$ follows from $B=\al/A$ and does not place any new constraints.
The solutions for \rf{Vp0case} are
\beq
\bal
B_0&=c_1-\frac{c_2}{r},\\
A&=\al+\frac{c_3}{r}-\frac{\ka\al {\bar V} r^2}{3}-\frac{\ka c_2^2}{2r^2},
\eal
\eeq
where the $c_n$'s are now distinct coefficients. 
Because $V'=0$, we must have that the argument of $V$ is constant, and thus
we can let $B_0^2/A=-\beta$, where $\beta$ is a non-negative constant, taking the value of one of the minima of the potential.  
For example, 
for the hypergeometric case, from equation \rf{hypv'}, $\be=3  \La^2$, 
and $\bar V$ is given by \rf{min3}.

The relation $B_0 = \pm \sqrt{-\be A}$ provides an additional constraint that must hold for the solution. 
Using this relation, 
we can combine the previous two equations to get the following
\beq
\bal
&\bigg{(}-\frac{\ka\al\be {\bar V}}{3}\bigg{)}r^2+\bigg{(}c_1^2+\al\be\bigg{)}+\bigg{(}c_3\be-2c_1c_2\bigg{)}\frac{1}{r}
\\
&\pt{sp}+\bigg{(}c_2^2-\frac{\ka\be c_2^2}{2}\bigg{)}\frac{1}{r^2}=0.
\eal
\eeq
As this expression must be true for all $r \neq 0$, 
we must have that the coefficients in front of each power of $r$ must be zero.  
If we consider the first coefficient of $r^2$, 
we must have either $\be=0$ or $V=0$ (discarding the possibility of $\al=0$ since that violates asymptotic flatness). 

In the former case, $\be=0$ implies $B_0=0$ (for finite $A$), 
which is a case we considered previously. 
Continuing,
we {\it also} get that $c_2=0$ from the coefficient in front of $1/r^2$ 
and $c_1=0$ from the coefficient of $r^0$. 
This leaves us with the solution
\beq
A=\al+\frac{c_3}{r}-\frac{\ka\al V_0r^2}{3}
\eeq
which is, as expected, 
identical to the first case \rf{SAdS}.

In the latter case, 
we have ${\bar V}=0$, and we will assume $B_0\neq0$ (and $\be \neq 0$). 
From the coefficient of $r^0$, we get that $c_1=\pm \sqrt{-\al\be}$. If we consider the coefficient of $1/r^2$, we see either $c_2=0$ or $\be=2/\ka$.
If $c_2=0$, then $c_3=0$ from the coefficient of $1/r$, and we arrive at the solution,
\beq
\bal
B_0&=\pm \sqrt{-\al\be},\\
A&=\al.
\label{flatsoln}
\eal
\eeq
This solution is Minkowski spacetime.
If $\be=2/\ka$, 
we see that $c_3=\ka c_2\sqrt{-\al\be}$, and we arrive at the solution
\beq
\bal
B_0&=\pm \left( \sqrt{-\al\be}-\frac{c_2}{r}\right), \\
A&=\al+\frac{\ka c_2\sqrt{-\al\be}}{r}-\frac{\ka c_2^2}{2r^2},
\eal
\eeq
which is of the RN form (we add a $\pm$ to the $c_2$ term in $B_0$
without loss of generality and to emphasize that $B_0^2=-\be A$).
To match with the standard form 
$A=-1 + 2GM/r -GQ^2/(4\pi r^2)$, 
$\al=-1$, 
$c_2 = \pm Q/ (4 \pi)$,
and charge and mass are related by $Q=\pm M/\sqrt{\be}$.
Since a nonzero $Q$ arises through the potential, 
this shows that a RN solution could arise in the absence of actual electric charge, 
as in Ref.\ \cite{Xu:2022frb}, 
and others.
Note that the constraint ${\bar V}=0$ can severely limit the possible form of the potential $V(X)$.

\begin{figure*}
\includegraphics[width=155mm]{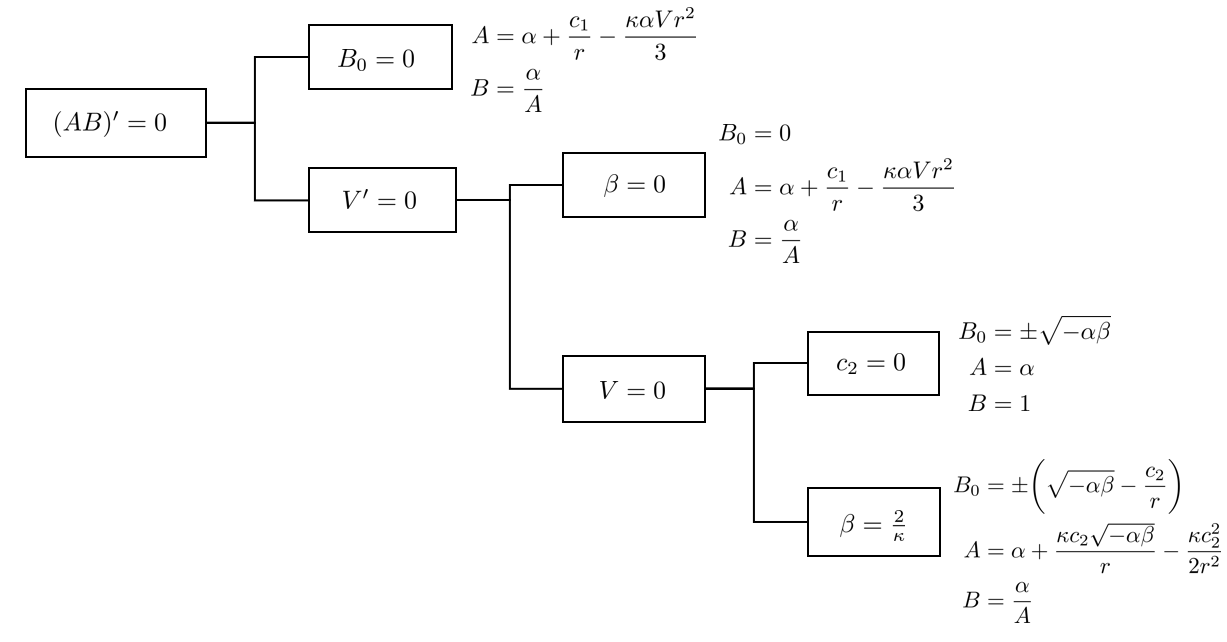}
 \caption{Flowchart showing the possible conditions and the resulting solutions of Eq.\ \rf{firstorder}, all under the assumption that $(AB)'=0$.}
 \label{flowchart}
\end{figure*}

The flowchart in Fig.\ \ref{flowchart} summarizes the special case solutions.
We leave it as an open problem to obtain exact solutions with assumptions other than $(AB)'=0$.

\section{Coupled Gravity-Vector Solutions, Part II}
\label{gravity-vector part II}

To proceed to analyze the landscape of possible solutions to the coupled ODEs in equations \rf{firstorder} 
when $V' \neq 0$, 
we change our coordinate system to one in which the behavior at any horizons, 
should they exist, 
is more tractable.
Thus we will use Eddington-Finkelstein (EF) coordinates
($v,r,\th, \ph$) \cite{mtw}, 
which are defined by the line element,
\beq
ds^2=-N(r) dv^2 + 2 M(r) dv dr + r^2 d\Om^2.
\label{EFle}
\eeq
The transformation between EF and  Schwarzschild coordinates ($t,r,\th,\ph$)
is accomplished by $dt=dv-(M/N) dr$, 
so that with \rf{ansatz2},
$B = M^2/N$,
and $A=-N$.
A primary utility of these coordinates for our purposes is that at a horizon the functions $N$, and $M$ do not diverge, rather $N(r_h)=0$ at a horizon $r_h$.
Also in these coordinates the vector field takes the form
$B_\mu=(B_v,B_r,0,0)$.

The field equations \rf{FEgravBB} evaluated with EF coordinates can be manipulated into the form,
\beq
\bal
B_v^{\prime \prime} + \frac{2}{r}B_v'
&= 2 V' \frac{M^2}{N} B_v + \ka V' \frac{M^2}{N^2}B_v^2 B'_v r,\\
M' &= \ka \frac{B_v^2}{N^2} M^3 V' r, \\
N' &= \frac {(M^2-N)}{r} -\frac {\ka}{2} (A'_v)^2 r
-\ka M^2 V r,
\label{EFeqns}
\eal
\eeq
where the argument of the potential becomes
$X=B_\mu B^\mu=-B_v^2/N$.
The simplification of these equations is due primarily to the identity 
$M B_v=-N B_r$, 
which follows from the $r$ component of the $B_\mu$ field equations (provided $V' \neq 0$).

Next we convert the equations to a dimensionless form for graphical, analytical and numerical study.
We define the dimensionless radial coordinate $x=r/r_h \ge 0$, 
where $r_h$ is a suitable horizon distance (e.g., $r_h=2GM$ in Schwarzschild geometry with standard boundary conditions).
Furthermore, 
we scale the vector field by a mass dimension, $f=B_v/\La$.
With $k=\ka \La^2$, 
we have,
\beq
\bal
f^{\prime \prime} + \frac{2}{x}f'
&= 2 \tV' \frac{M^2}{N} f + k \tV' \frac{M^2}{N^2}f^2 f' x,\\
M' &= \ka \frac{f^2}{N^2} M^3 \tV' x, \\
N' &= \frac {(M^2-N)}{x} -\frac {\ka}{2} (f')^2 x
-\ka M^2 \tV x,
\label{EFeqnsdim}
\eal
\eeq
where the dimensionless potentials are $\tV = (r_h^2/\La^2 ) V$
and $\tV' = r_h^2 V'$.
For convenience, 
we define dimensionless ``strengths" for the potentials as follows: $\tilde{\mu}^2 = \mu^2 r_h^2$,
$\tilde{\la} = \la r_h^2 \La^2$,
and $\tilde{g}=g r_h^2 \La^2$.

The system of equations in \rf{EFeqnsdim} are second order nonautonomous (an explicit $x$ dependence) ordinary differential equations.
For $x > 0$
and $N \neq 0$,
the solutions are unique once initial values for $f$, $f'$, $M$, and $N$ are given at some $x_0 \neq 0$
\cite{sanchez1979ordinary}.
The system \rf{EFeqnsdim} can 
be made first order by using $z=f'$ to yield a first order equation.
It is still technically nonautonomous so many of the ODE techniques for characterizing solutions based on integral curves for example will not apply (at least not in a straightforward manner).

The arguments for the global behavior of the function $f$ made in the previous section \ref{analyticargument} are no longer straightforward due to the extra term proportional to the gravity coupling $k$ on the right-hand side of the $f$ equation in \rf{EFeqnsdim}.
This term is proportional to $f'$ and hence it is difficult to make arguments about its sign.
However, 
one can discern the behavior of $M$ more easily than the $N$ or $f$.
Assuming $M>0$,
all the terms on the right-hand side are positive save $\tV'$.
Thus the equation for $M'$ tells us that
\beq
\bal
M' &< 0, \tV' <0,\\
M' &> 0, \tV' >0.
\label{}
\eal
\eeq
One can then reason what the global behavior of $M$ will be using the forms of the different potential choices, 
at least under certain conditions on $f$ and $N$.

\subsection{Solutions near horizon and the asymptotic region}
\label{analytic}

For this part, 
we explore solutions that can be expanded near various points of interest.
Namely, 
we wish to study solutions in the neighborhood of a horizon, 
and also in the $x \rightarrow \infty$ asymptotic limit.
For this part, 
we will assume the functions $N$, $M$, and $f$
are analytic near the horizon and asymptotically 
(similar to past works \cite{Bardeen:1999px,Eling:2006ec}).

\subsubsection{Horizon}
\label{Horizon}

We consider first the assumption of a regular horizon.  
In EF coordinates this means $N(1)=0$
and $N(x),M(x),f(x)$ are all analytic functions in the neighborhood of $x=1$ ($r=r_h$).
We form series expansions for these functions in terms of $\de x=x-1$:
\beq
\bal
f(x) &= \sum_{j=0}^{m} a_f(j) \de x^j,
\\
M(x) &= \sum_{j=0}^{m} a_M(j) \de x^j,
\\
N(x) &= \sum_{j=1}^{m} a_N(j) \de x^j,
\label{horizonexp}
\eal
\eeq
where $m$ is the desired truncation.
By construction, this expansion must have $a_N(0)=0$.
The expansions \rf{horizonexp}
are inserted into the ODEs \rf{EFeqnsdim}, 
and the coefficients are fixed order by order in $\de x^j$ up to the desired term truncation.
Running through this calculation we find computation time is saved if we carry out this calculation to obtain an expansion to order $\de x^5$ or $\de x^6$, 
which, 
due to the derivatives and small denominator terms, 
requires at least $m=8$ in equation \rf{horizonexp}.

We first note the form of the expansion for the RN case when $\tV=0$.
In this case of course, 
the exact solution is known \rf{RNsoln}.
Expanding the around the horizon $x=1$ 
(so that $x=1+\de x$), 
we have
\beq
\bal
f(x) &= a_f (0) + a_f (1) (\de x - \de x^2  +...), \\
M(x) &= 1, \\
N(x) &= (1 - \tfrac 12 a_f(1)^2 )\de x
+(k a_f(1)^2 -1) \de x^2 +...,
\label{RNhorizon}
\eal
\eeq
where the solutions depend only on the horizon values $a_f (0)$ and $a_f(1)$.
Clearly, 
the type of potential $\tV$ chosen influences the relations among the coefficients $a_f(j)$, 
$a_M(j)$, 
and $a_N(j)$.
For any choice $\tV \neq 0$ of potential, 
we find $a_f(0)=0$ and 
the coefficient $a_f(1)$ is the only free parameter left, 
fixed by the value of $df/dx$ at the horizon.
To capture the dependencies of the coefficients, we show them explicitly for the first three terms in the series in Tables \ref{horizon1} and \ref{horizon2}.

\begingroup
% Default value: 6pt
\renewcommand{\arraystretch}{2} % Default value: 1
\begin{table*}[!h]
 \caption{First three terms in the horizon analytic expansion \rf{horizonexp} for the RN and massive potential cases.  We take $k=1$, $a_M(0)=1$.  Also, $\al_1=2-\af1^2$, $\al_2=1-\af1^2$ and we abbreviate $\af1=a_f(1)$.}
 \begin{tabular}{ |p{0.6cm}||p{1.2cm}|p{1.2cm}|p{1.2cm}||p{2.3cm}|p{4.2cm}|p{2.8cm}|}
 \hline
 \multicolumn{7}{|c|}{Horizon expansion coefficients} \\
 \hline
  & \multicolumn{3}{|c||}{RN} & \multicolumn{3}{|c|}{Massive} \\
 \hline
 j  & $a_f(j)$ & $a_M(j)$ & $a_N(j)$  & 
 $a_f(j)$ & $a_M(j)$ & $a_N(j)$ \\
\hline
0  & $\af0$ & $1$ & $0$  & 
$0$ & $1$ & $0$ \\
\hline
1 & $\af1$ & $0$ & $\tfrac 12 \al_1$ &
$\af1$ & $2 \t{\mu}^2 \af1^2/\al_1^2$ & $\tfrac 12 \al_1$ \\
\hline
2  & $-\af1$ & $0$ & $ -\al_2 $ &
%%%Massive below%%%%%%
$\af1 \left(\frac {2 \t{\mu}^2}{\al_1^2}-1 \right) $ & 
$\frac {\t{\mu}^2 \af1^2 [ 4-8\af1^2 +3 \af1^4 + 4\t{\mu}^2 (1+\af1^2)]}{\al_1^4} $ & 
$\frac { [(6+\t{\mu}^2)\af1^2 -2 \af1^4-4]}{2\al_1}$ 
\\
\hline
\end{tabular}
 \label{horizon1}
\end{table*}
\endgroup

\begingroup
\renewcommand{\arraystretch}{2} % Default value: 1
\begin{table*}[!h]
 \caption{The first three terms in the horizon analytic expansion \rf{horizonexp} for the quadratic and two sample hypergeometric potentials.  
 We take $k=1$, $a_M(0)=1$.
 Also, we use abbreviations:
$\be_1=\t{g}+e^3 (2 \t{g}-\al_1)$, 
$\be_2=\t{g}+2e^3 (\t{g}-1)$,
$\be_3=2\t{g} + 3e^2 (2\t{g}-\al_1)$,
and $\be_4=\t{g}+ 3e^2 (\t{g}-1)$.
 }
\begin{tabular}{ |p{0.5cm}||p{3.8cm}|p{7.6cm}|p{4.5cm}|}%ten columns
 \hline
 \multicolumn{4}{|c|}{More horizon expansion coefficients} \\
 \hline
   & \multicolumn{3}{|c|}{Quadratic}   \\
 \hline
 j  & $a_f(j)$ & $a_M(j)$ & $a_N(j)$  \\
\hline
0  & 0 & 1 & 0  \\
\hline
1  &  $\af1 $ & $\frac {4 \t{\la} \af1^2}{(\t{\la}-\al_1)^2}$ & $\tfrac 12 (\al_1-\t{\la})$ \\
\hline
2  &  $-\af1 \frac {[3\t{\la}^2 -2 \t{\la}(2+\al_1) +\al_1^2]}{(\t{\la}-\al_1)^2} $ & 
$\frac {2 \t{\la} \af1^2 [4-5\t{\la}^2 -12 \af1^2 +5 \af1^4 + 4\t{\la} (2+3\af1^2)]}{(\t{\la} - \al_1)^4} $ & 
$\frac { (2-\t{\la}-3\af1^2+\af1^4)}{\t{\la} -\al_1}$ 
\\
\hline
%%%% M(3,2,z)
  & \multicolumn{3}{|c|}{$M(3,2,z)$}   \\
 \hline
 j  & $a_f(j)$ & $a_M(j)$ & $a_N(j)$  \\
\hline
0  & $0$ & $1$ & $0$  \\
\hline
1  & $\af1$ & $6e^6 \t{g}  \af1^2/\be_1^2$ & $-\be_1/(2 e^3)$ \\
\hline
2  & $-\af1 \left(1+ \frac{3e^3 \t{g} \be_2}{\be_1^2} \right)$ & 
$ \frac{e^6 \t{g} \af1^2 [ 
-3 \be_2 (\t{g}+e^3 (2+8\t{g})) 
+2e^3 (7\t{g} +4e^3 (8\t{g}-5))\af1^2
+17e^6 \af1^4 ]}{\be_1^4}$ & 
$\frac { -2\t{g} \al_2 +e^3 (4-4\t{g} +(\t{g}-6)\af1^2 + 2\af1^4 ) }{2\be_1}$  \\
\hline
%%%% M(4,2,X/\La^2)
& \multicolumn{3}{|c|}{$M(4,2,z)$}\\
\hline
 j  & $a_f(j)$ & $a_M(j)$ & $a_N(j)$  \\
\hline
0  & $0$ & $1$ & $0$  \\
\hline
1  & $\af1$ & $72e^4 \t{g} \af1^2/\be_3^2$ & $-\be_3/6e^2$ \\
\hline
2 & $-\af1 \left( 1+ \frac{24e^2 g \be_4}{\be_3^2}  \right)$ &  
$ \frac{36 e^4 \t{g} \af1^2
[
57e^4 \af1^4 -4 \be_4 (\t{g}+3e^2 (1+5\t{g}) )
+4e^2 (8\t{g} +e^2 (60\t{g}-33) ) \af1^2
]}{\be_3^4}$ & 
$\frac {- 2 \t{g} \al_2  +3 e^2 (2-2\t{g} -3\af1^2 +\af1^4) }{\be_3}$  \\
\hline
\end{tabular}
\label{horizon2}
\end{table*}
\endgroup

From the table one can see some general features.
As a guide, 
we focus on the horizon values that yield an RN solution that asymptotes to a constant value.
For this behavior, 
from the RN entry in Table \ref{horizon1}, 
we would need $\al_1>0$, so that $N$ is increasing from $0$ and we need to ensure the second derivative is negative, or $-\al_2<0$.
These two criteria can be achieved if 
$\af1^2 <1$.
Given these criteria, 
we examine the behavior implied by the four potential cases in Tables \ref{horizon1} and \ref{horizon2}.
For the massive potential, 
the behavior does depend on the size of $\t{\mu}$.
Looking at the behavior of $N$, 
we see that $dN/dx>0$ if $\al_1>0$.
If $\af1=0$, the solution appears to reduce to the Schwarzschild behavior with the vector field vanishing (this is consistent with known results).
If $\af1^2 <1$ (but nonzero), 
then one can show that $d^2N/dx^2>0$ for all nonzero $\t{\mu}$ which can yield nonasymptotically flat behavior.
The behavior of $f(x)$ is determined by the sign of $\af1$, 
while $M$ will always be initially increasing with concavity of varying sign depending on $\t{\mu}$.
%%check if \af1^2>1

Moving on to the quadratic potential case, 
if $\al_1>0$, 
the sign of $dN/dx$ will depend sensitively on the value of $\t{\la}$.
The second derivative of $N$ also depends sensitively on $\t{\la}$.
For instance, 
if $\t{\la}>>1$, 
then $N$ will decrease and $N''<0$, 
which does not resemble a solution closely following the RN solution.
If $\t{\la}<<1$, 
then $dN/dx>0$ and $N''<0$ or $N''>0$, 
depending on the value of $\af1^2$.

For the hypergeometric potentials it is more challenging to discern which choices of the coupling $\t{g}$
and the value $\af1$
will yield $dN/dx>0$
and $N''<0$ as in the RN case.
Some studies of the dependence of $N'$ and $N''$
on these parameters
for the $M(3,2,z)$ case and the $M(4,2,z)$,
reveal that for a range of values described approximately by $\t{g}^2+\af1^2 <1$ (with $\af1>0$ and $\t{g}>0$), 
then $N'>0$ and $N''<0$ at the horizon.
So it is possible for a solution to initiate behavior similar to the RN case.
However, 
the solution for $M$ in this case has $M'>0$ and $M''>0$, indicating a possible divergence.

These remarks only apply for $x$ near the horizon ($\de x<<1$). 
Ultimately, 
the ``exact" solution will show whether solutions mimic the RN behavior or not as $x$ increases.
The next step is to use the near-horizon behavior to generate initial conditions for precise numerical solutions.
This is carried out in Mathematica \cite{Mathematica}.
We also carry out the horizon expansion up to fifth or six order in $\de x$.
The higher coefficients $a_f(j)$, $a_M(j)$, and $a_N(j)$ for $j>2$ are much too lengthy to record here but are stored in Mathematica files.
For a numerical solver, 
the differential equations \rf{EFeqnsdim} are singular at $N=0$ ($x=1$) so we step away from this point slightly ($x=1.05$) and use initial conditions from the horizon expansion at that point.

In the plots to follow, 
we show both the horizon expansion, 
and the numerical results together for several choices of potential, 
and sample initial values at the horizon.
No approximation is made for the numerical solutions; the full equations are used.
The numerical scheme, 
employing ``NDSolveValue" in Mathematica, 
uses a combination of varying order Runge-Kutta techniques 
designed to carefully approach stiff or singular points.
One can see where the horizon expansion breaks down in these plots.
A common feature we discern is that for various choices of potential, 
the full numerical code appears to carve out quite a different path from the initial trends 
indicated by the horizon expansion.

\begin{figure*}
  \centering
     \subfloat[The massive vector potential case.  The series solution is truncated to $\de x^5$.  The initial value is $f'(1)=-1/3$ and ${\tilde \mu}=1$ at the horizon.]{
         \includegraphics[width=0.48\textwidth]{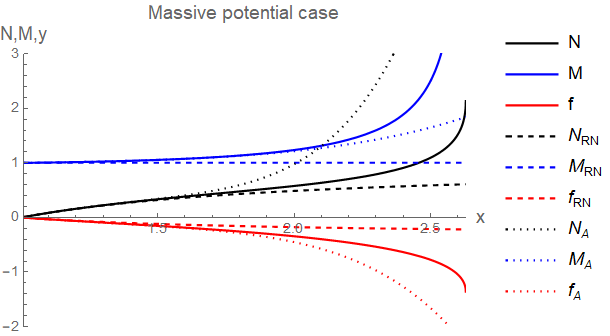}
         \label{horizonM}}
     \hfill
     \subfloat[The quadratic potential case. The series solution is truncated to $\de x^5$.
         The initial value is $f'(1)=-1/5$ and ${\tilde \la}=1/5$ at the horizon.]{
         \includegraphics[width=0.48\textwidth]{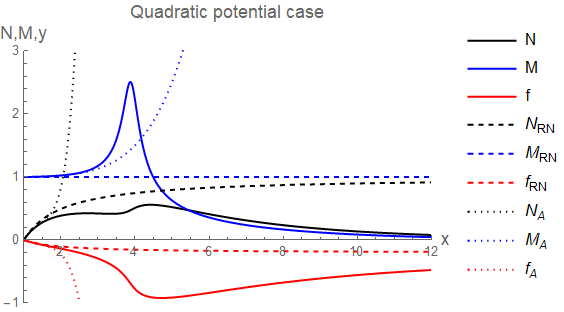}
         \label{horizonQ}} 
         \\
     \subfloat[The hypergoemetric potential $M(3,2,z)$ case.  The series solution is truncated to $\de x^6$.
         The initial value used is $f'(1)=-1/10$, while  and ${\tilde g}=1/10$.]{
         \includegraphics[width=0.48\textwidth]{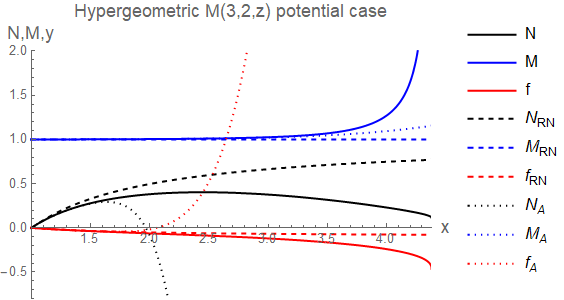}
         \label{horizonM3}} 
      \hfill
      \subfloat[The hypergeometric potential $M(4,2,z)$ case.  The series solution is truncated to $\de x^5$.
         The initial value used is $f'(1)=-1/9$, while  and ${\tilde g}=1/8$.  
         The peak for the $M$ curve extends to $M \approx 7.5$.]{
         \includegraphics[width=0.48\textwidth]{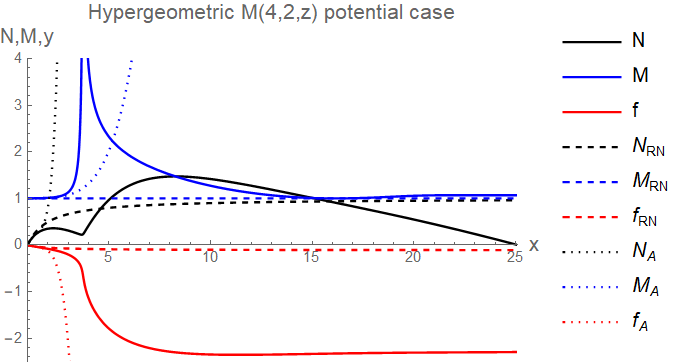}
         \label{horizonM4}} 
      \caption{Numerical solutions using the horizon expansion \rf{horizonexp} to generate initial values near $x=1$.  The analytical results with the series solution are labeled with subscript $A$.  
        The dashed lines are the RN case while the dotted lines show the analytical series from equation \rf{horizonexp}.  The connection with conventional Schwarzschild coordinates is $g_{tt}=A=-N$ and $z=X/\La^2$.}
       \label{horizonplots}
\end{figure*}
     
     % \begin{subfigure}[b]{0.49\textwidth}
     %     \centering
     %     \includegraphics[width=\textwidth]{QuadraticHorizon.png}
     %     \caption{The quadratic potential case. The series solution is truncated to $\de x^5$.
     %     The initial value is $f'(1)=-1/5$ and ${\tilde \la}=1/5$ at the horizon.}
     %     \label{horizonQ}
     % \end{subfigure}
     % \begin{subfigure}[b]{0.49\textwidth}
     %     \centering
     %     \includegraphics[width=\textwidth]{M3Horizon.png}
     %     \caption{The hypergoemetric potential $M(3,2,z)$ case.  The series solution is truncated to $\de x^6$.
     %     The initial value used is $f'(1)=-1/10$, while  and ${\tilde g}=1/10$.}
     %     \label{horizonM3}
     % \end{subfigure}
    % \hfill
    %  \begin{subfigure}[b]{0.49\textwidth}
    %      \centering
    %      \includegraphics[width=\textwidth]{M4Horizon.png}
    %      \caption{The hypergeometric potential $M(4,2,z)$ case.  The series solution is truncated to $\de x^5$.
    %      The initial value used is $f'(1)=-1/9$, while  and ${\tilde g}=1/8$.  
    %      The peak for the $M$ curve extends to $M \approx 7.5$.}
    %      \label{horizonM4}
     % \end{subfigure}
        % \caption{Numerical solutions using the horizon expansion \rf{horizonexp} to generate initial values near $x=1$.  
        % The analytical results with the series solution are labeled with subscript $A$.  
        % The dashed lines are the RN case while the dotted lines show the analytical series from equation \rf{horizonexp}.  The connection with conventional Schwarzschild coordinates is $g_{tt}=A=-N$ and $z=X/\La^2$.}
        % \label{horizonplots}

The first plot Fig.\ \ref{horizonM}, is the massive vector potential case.
The plot shows that when there is a horizon, the fields $N$, $M$, $f$ show diverging behavior - 
which occurs in the numerical simulation.\
This implies that there is either another horizon or a physical singularity in the spacetime.
The situation is resolved by checking the curvature 
in Section \ref{curvature}, 
which shows that this solution, which starts from a horizon, 
leads to a physical singularity.
Thus the solution is not asymptotically flat; consistent with the Bekenstein result.

Consider next the quadratic potential case
in the second plot Fig.\ \ref{horizonQ}.
This plot shows that when the solution starts from a horizon at $x=1$, 
there is an interesting pulse like behavior in $M$
at about $r \approx 4r_h$, followed by a decrease to zero for $N$ and $M$.
Again, 
we can ask the nature of the singularity as $N \rightarrow 0$.
The answer provided by the curvature is that as $x\rightarrow \infty$, 
scalar curvature quantities appear to diverge, suggesting a singularity.
%recheck???  smooth divergence???

Now we consider the first hypergeometric potential $M(3,2,z)$.
The behavior indicated by the sample in Fig.\ \ref{horizonM3} suggests a regular solution 
for the functions $f$, $M$, and $N$ until one reaches $x=4.38$, 
where the numerical code fails (for the given initial conditions).
We have analyzed this region (near $x=4.38$) with calculation of curvature scalars 
and we find disagreement (see the curvature section below).
For example, 
the Ricci scalar we find must be finite at $x=4.38$, 
in disagreement with the numerical calculations.
This behavior is consistent when one varies the initial values of $f'$ and the coupling ${\tilde g}$; 
different choices merely move the stiff point to larger or shorter $x$.
Thus it remains an open question to decipher the behavior at these points, 
whether they are coordinate singularities or physical ones.

For the case of $M(4,2,z)$, 
the results of the sample in Fig.\ \ref{horizonM4} show an interesting pulse-like behavior 
similar to the quadratic potential case.
Around $x=25$, 
the function $N$ hits zero again, 
indicated another possible horizon.
Indeed, 
calculations of the curvature show that it is finite at $x=25$.
Thus we can say that this case contains RN-like behavior with two horizons.
The pulse-like behavior persists when one changes the horizon values of $f'(1)$ and $\t{g}$, 
as discussed below in Sec.\ \ref{curvature}.

Not shown is the case for $M(5,2,z)$, 
for which we find that the behavior is similar to the case $M(4,2,z)$.
The calculation of the curvature also confirms that when $N$ approaches zero, 
the curvature is finite as indicated from curvature invariants.
For both $n=4,5$ hypergeometric potentials it remains an open problem to map out the spacetime beyond 
this value of $x$ for which one gets horizon behavior. 

\subsubsection{Asymptotic region}
\label{asymptotic}

Here we explore solutions in the asymptotic region.
We first consider an analytical solution as $x\rightarrow \infty$ by using the variable $u=1/x$ from Section \ref{spherical nonlinear}. 
Thus, we consider an analytic series expansion around $u=0$. 
As in the horizon analysis in Sec.\ \ref{Horizon} we compare with numerical solutions for the full domain of $u$ save any singular points.
In Sec.\ \ref{non-analytic} we examine approximate solutions in the asymptotic region that are not analytic at $u=0$.

The analytic expansion for the solutions takes the form,
\beq
\bal
f(u) &= \sum_{j=0}^{m} c_f(j) u^j,
\\
M(u) &= \sum_{j=0}^{m} c_M(j) u^j,
\\
N(u) &= \sum_{j=0}^{m} c_N(j) u^j.
\label{asymptoticexp}
\eal
\eeq
The coefficients $c_f(j)$, $c_M(j)$, 
and $c_N(j)$ are to be determined by insertion into 
the equations \rf{EFeqnsdim} re-expressed in terms of $u$.
In particular the equations take the form, 
for this case,
\beq
\bal
\frac {d^2f}{du^2} &= 2 f \frac{M^2 \tV'}{N u^4} 
-k f^2 \frac{M^2 \tV'}{N^2 u^3} \frac{df}{du},\\
\frac {dM}{du} &= -k f^2 \frac{M^3 \tV'}{N^2 u^3},\\
\frac{dN}{du} &= \frac {N-M^2}{u} +\frac{k}{2} u \left( \frac {df}{du} \right)^2
+ k \frac {M^2}{u^3} \tV.
\label{ueqns}
\eal
\eeq

We then solve order by order in $u$.
Note that this implies that the coefficients in the expansion must take values to cancel terms with inverse powers of $u$, starting with 
$1/u^4$.
We include tables with the first three terms in the series for the different choices of potential.

\begingroup
\renewcommand{\arraystretch}{2} 
\begin{table*}
 \caption{First 3 terms in the asymptotic analytic expansion \rf{asymptoticexp} for the RN, and quadratic potential cases.  We set $k=1$ for simplicity. 
 The massive potential solution is identical to the $RN$ case but with $\cf0=0=\cf1$.}
\begin{tabular}{ |p{0.5cm}||p{2cm}|p{2cm}|p{2cm}||p{2cm}|p{2cm}|p{2cm}|}
%7 columns
 \hline
 \multicolumn{7}{|c|}{Asymptotic ($1/x=u \rightarrow 0$) expansion coefficients} \\
 \hline
  & \multicolumn{3}{|c||}{RN} & \multicolumn{3}{|c|}{Quadratic}\\
 \hline
 j  & $c_f(j)$ & $c_M(j)$ & $c_N(j)$  & $c_f(j)$ & $c_M(j)$ & $c_N(j)$  \\
\hline
0 & $\cf0$ & $\sqrt{\cN0}$ & $\cN0$  
& $\cf0$ & $-\cf0$ & $\cf0^2$  \\
\hline
1 & $\cf1$ & $0$ & $\cN1$
& $\cf1$ & $0$ & $2 \cf0 \cf1$  \\
\hline
2 & $0$ & $0$ & $k \cf1^2/2$
& $-\cf1^2/\cf0$ & $k\cf1^2/(2\cf0 )$ & $0$  \\
\hline
3 & $0$ & $0$ & $0$
& $\frac{(6+k)\cf1^3}{2\cf0^2}$ & $-4k\cf1^3/\cf0^2$ & $k\cf1^3/\cf0$  \\
\hline
\end{tabular}

 \label{asymptotic1}
\end{table*}
\endgroup

\begingroup
% Default value: 6pt
\renewcommand{\arraystretch}{2} % Default value: 1

\begin{table*}
 \caption{First 3 terms in the asymptotic analytic expansion \rf{asymptoticexp} for the hypergeometric functions for $n=3$ and $n=4$.}
\begin{tabular}{ |p{0.5cm}||p{2.2cm}|p{2.2cm}|p{2.2cm}||p{2.2cm}|p{2.2cm}|p{2.2cm}|}
%7 columns
\hline
\multicolumn{7}{|c|}{Asymptotic ($1/x=u \rightarrow 0$) expansion coefficients} \\
\hline
  & \multicolumn{3}{|c||}{$M(3,2,z)$} & \multicolumn{3}{|c|}{$M(4,2,z)$}\\
\hline
j  & $c_f(j)$ & $c_M(j)$ & $c_N(j)$  & $c_f(j)$ & $c_M(j)$ & $c_N(j)$  \\
\hline
0 & $\cf0$ & $-\cf0/\sqrt{3}$ & $\cf0^2/3$  
& $\cf0$ & $-\cf0/\sqrt{2}$ & $\cf0^2/2$  \\
\hline
1 & $\cf1$ & $0$ & $\tfrac 23 \cf0 \cf1$
& $\cf1$ & $0$ & $ \cf0 \cf1$  \\
\hline
2 & $-\cf1^2/\cf0$ & $-\frac{\sqrt{3}k\cf1^2}{2\cf0}$ & $0$
& $-\cf1^2/\cf0$ & $-k\cf1^2/(\sqrt{2}\cf0 )$ & $0$  \\
\hline
3 & $\frac{3(2+k)\cf1^3}{2\cf0^2}$ & $4\sqrt{3}k\cf1^3/(\cf0^2)$ & $k\cf1^3/\cf0$ 
& $(3+k)\cf1^3/(\cf0^2)$ & $4\sqrt{2}\cf1^3/(\cf0^2)$ & $k\cf1^3/\cf0$  \\
\hline
\end{tabular}
 \label{asymptotic2}
\end{table*}
\endgroup

Tables \ref{asymptotic1} and \ref{asymptotic2} display some general features 
for how the functions behave in the asymptotic region.
Firstly, 
the initial data needed to determine the behavior rely only on 
$f$ and $df/du$ at $u=0$.
For the non-RN cases of the quadratic and hypergeometric potentials, 
the signs of $\cf0$ and $\cf1$ are critical to the behavior of the solutions for increasing $u$, 
and thus decreasing $x$.
Regarding the function $N$, 
if $\cf1 \cf0>0$ then $N$ will be increasing up to order $u^3$, 
and the order $u^2$ term vanishes.
This behavior resembles that of a naked singularity.
Indeed this is borne out by studying numerical solutions.
On the other hand if $\cf1 \cf0<0$, then $N$ will start off decreasing.
For the latter condition, 
this is a good starting constraint on the initial values in order to try to find solutions that approach a horizon, 
like the conventional RN case.

The function $M$ has a different behavior.  
The $O(u)$ term vanishes for all potentials.
The second and third order terms
can be positive or negative depending on choices of $\cf0$ and $\cf1$.
This was to be expected given the $M' \sim M^3$ structure in equation \rf{ueqns}.

Another striking feature for the potentials considered with a nonzero minimum 
for the vector field (quadratic and hypergeometric choices) emerges.
Glancing at Tables \ref{asymptotic1} and \ref{asymptotic2}, 
in the columns for $M$ and $N$ coefficients, 
reveals what happens when the choice of $\cf1=0$ for $u=0$ is made.
For this choice, 
{\it all of the} coefficients beyond the $j=0$ row vanish; as they are all proportional to powers of $\cf1$.
This holds to higher order in the expansion \rf{asymptoticexp} and for numerical solutions beyond this below.
Thus, 
the naive suggestion that the vector field should approach a constant value 
with vanishing derivative at $x\rightarrow \infty$ 
reduces all solutions to Minkowski spacetime!
Of course, 
this is all assuming that the solution is analytic in $u$
as $u \rightarrow 0$.
%can this be gleaned from the ODEs??

We can also confirm the Bekenstein result for the massive vector field, 
as the exact solution for this case emerges from the asymptotic expansion.
Specifically, 
the solutions for the massive potential require that $\cf0=0=\cf1$,
so this case reduces to the Schwarzschild solution with zero vector field.
%check with papers - other papers seem to allow naked singularities for Proca potential, but that would appear to be ruled out here????
% this assumes analytic asymptotic...

\begin{figure*}
  \centering
   \subfloat[The quadratic potential case with asymptotic values for the functions of order unity.  The initial values are taken to be $f(0)=-1$, $f'(0)=1/2$ and we set ${\tilde \la}=1/5$.]{
         \includegraphics[width=0.48\textwidth]{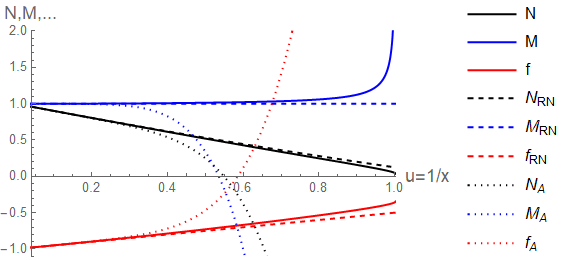}
         \label{asympQ}}
     \hfill
   \subfloat[A second case of the quadratic potential with initial values of $f(0)=-1$, $f'(0)=-1/5$, and ${\tilde \la}=1$.  This behavior is consistent with a naked singularity.]{
         \includegraphics[width=0.48\textwidth]{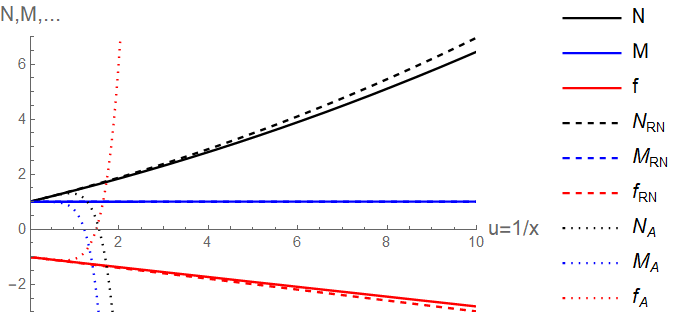}
         \label{asympQ2}}
     \\
   \subfloat[The hypergeometric potential $M(3,2,z)$ case.  
         The asymptotic values used are $f(0)=-\sqrt{3}$ and $f'(0)=1/\sqrt{3}$ while ${\tilde g}=1$ and $k=1/2$.]{
         \includegraphics[width=0.48\textwidth]{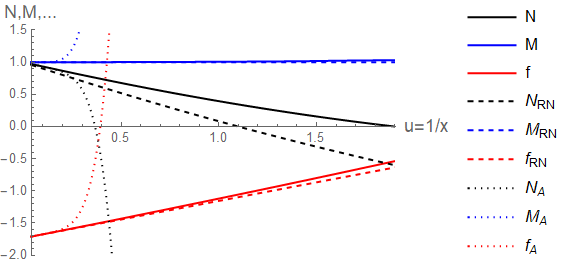}
         \label{asympM3}}
     \hfill
   \subfloat[The hypergeometric potential $M(4,2,z)$ case.  
         The initial values are $f(0)=-\sqrt{2}$ and $f'(0)=1/\sqrt{2}$, while $k=2/3$ and ${\tilde g}=1$.]{
         \includegraphics[width=0.48\textwidth]{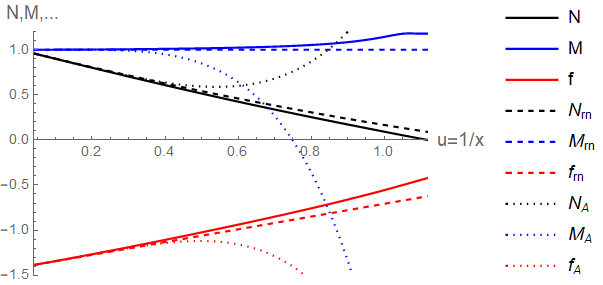}
         \label{asympM4}}
  \caption{Numerical solutions using the asymptotic expansion \rf{asymptoticexp} to generate initial values near $u=0$.  The analytical results with the series expansion are labeled with subscript $A$.  The dashed lines are the RN case while the dotted lines show the analytical series from \rf{asymptoticexp}; the latter includes terms up to $u^6-u^7$.  The connection with conventional Schwarzschild coordinates is $g_{tt}=A=-N$.}
    \label{asymptoticplots}
\end{figure*}

We next show plots (Fig.\ \ref{asymptoticplots}) for these potential choices,
as in the previous section, 
assuming $\cf1 \neq 0$.
We use the asymptotic expansion to generate boundary conditions for the full numerical solver.
Specifically, 
since the numerical solver has difficulty handling the point $u=0$, 
we start with initial values of $f$ and $df/du$ at $u=1/20$, 
the latter generated from the analytical expansion \rf{asymptoticexp} truncated to order $u^5$ or $u^6$ (computing time increases with the order and varies with the type of potential).
One can see this in the plots where the numerical solution and the analytical solution match near $u=0$.

     % \begin{subfigure}[b]{0.49\textwidth}
     %   \centering
     %     \includegraphics[width=\textwidth]{Qasv2.png}
     %     \caption{The quadratic potential case with asymptotic values for the functions of order unity.  
     %     The initial values are taken to be $f(0)=-1$, $f'(0)=1/2$ and we set ${\tilde \la}=1/5$.}
     %     \label{asympQ}
     % \end{subfigure}
     % \hfill
     % \begin{subfigure}[b]{0.49\textwidth}
     %     \centering
     %     \includegraphics[width=\textwidth]{QasNS.png}%naked sing
     %     \caption{A second case of the quadratic potential with initial values of $f(0)=-1$, $f'(0)=-1/5$, and ${\tilde \la}=1$.  This behavior is consistent with a naked singularity.}
     %     \label{asympQ2}
     % \end{subfigure}
     % \begin{subfigure}[b]{0.49\textwidth}
     %     \centering
     %     \includegraphics[width=\textwidth]{M3as.png}
     %     \caption{The hypergeometric potential $M(3,2,z)$ case.  
     %     The asymptotic values used are $f(0)=-\sqrt{3}$ and $f'(0)=1/\sqrt{3}$ while ${\tilde g}=1$ and $k=1/2$.}
     %     \label{asympM3}
     % \end{subfigure}
    % \hfill
    %  \begin{subfigure}[b]{0.49\textwidth}
    %      \centering
    %     \includegraphics[width=\textwidth]{M4as.png}
    %      \caption{The hypergeometric potential $M(4,2,z)$ case.  
    %      The initial values are $f(0)=-\sqrt{2}$ and $f'(0)=1/\sqrt{2}$, while $k=2/3$ and ${\tilde g}=1$.  }
    %      \label{asympM4}
    %  \end{subfigure}

Several features emerge from the figures.
Asymptotic analytic solutions appear to have the feature that they can either approach a horizon, 
become a naked singularity, 
or behave in a hybrid manner.
This behavior depends critically on the asymptotic values.

For the quadratic potential, 
shown in Figs.\ \ref{asympQ} and \ref{asympQ2}, 
two choices of initial values are made, 
one which yields decreasing behavior of $N$, 
the other increasing behavior.
For the former, 
while $N$ appears headed to zero, 
$M$ increases without limit, 
in what appears to be a singular or stiff point
in the numerical simulation.
From curvature analysis, 
the point $u \approx 1.002$ appears to be a physical singularity of a different type from conventional ones (because $N\rightarrow 0$ and $M\rightarrow \infty$).
We hesitate to call it a naked singularity because it may be partially hidden since $N \rightarrow 0$.
For the latter case in Fig.\ \ref{asympQ2},
qualitatively different boundary conditions yield a clear naked singularity.

For the hypergeometric cases, 
boundary conditions can also be chosen to yield naked singularity solutions like the quadratic case.
We focus on the case where initial values are chosen such that $N \rightarrow 0$.
Again, we wish to see if these solutions naturally contain a horizon.

For the $M(3,2,z)$ case, 
we display a plot for which the asymptotic values will yield $dN/du <0$ initially.
As shown in Fig.\ \ref{asympM3}, 
the solution indeed appears to have $N$ reach $0$.
However, 
at this point the numerical code breaks down.
Studies of curvature scalars show a smoothly increasing Kretschmann curvature scalar, 
revealing that
there appears to be a singularity at $u \approx 1.90$.
The smoothness gives an indicator of the reliability of the numerical solver \cite{tong09}.
This is similar to the case encountered for the quadratic potential.

For the $M(4,2,z)$ case, 
we again display a plot for which the asymptotic values will yield $dN/du <0$ initially.
As shown in Fig.\ \ref{asympM4}, 
the solution appears to have $N$ reach $0$.
At this point, 
the numerical code breaks down;
the curvature scalars show a smoothly increasing Kretschmann scalar, again 
indicating a singularity at $u \approx 1.12$.

As in the horizon expansion case in Section \ref{Horizon}, 
it would be of interest to further study these cases over a range of values for the initial data.
We leave this for future work.

\subsubsection{Curvature}
\label{curvature}

Invariant quantities can show unequivocally if
singularities exist in the solutions.
We first examine the Kretschmann scalar $K=R_{\mu\nu\ka\la} R^{\mu\nu\ka\la}$, 
which is one among the many curvature invariants in four-dimensional Riemann spaces 
\cite{Carminati:1991ddy,Overduin:2020aiq}.
%also cite 1997 paper and others?
In EF coordinates $K$ takes the form,
\beq
\bal
K&= \frac {1}{r^4 M^6} \big[ M^{\prime 2} (8 r^2 N^2 +r^4 N^{\prime 2}) - 2 r^2 M M' N' (4N 
\\
& \pt{=}+ r^2 N'')
+M^2
( 4(M^2-N)^2 +4 r^2 N^{\prime 2} + r^4 (N'')^2 )\big]\label{Kretschmann}
\eal
\eeq
The dimensionless version, 
dependent on $x=r/r_h$, follows from a mere scaling of the expression with $r_h^4$.

\begin{figure*}
  \centering
     \subfloat[A plot of the Kretschmann scalar $K=R^{\mu\nu\ka\la}R_{\mu\nu\ka\la}$ for the case of plot \ref{horizonM4}, with small parameter variations.  The curvature near $x=1$ is actually finite (not shown).  We chose the vertical range to exhibit the interesting bump feature.  The curves are distinguished by the values of $(\t{g},k,f'(1))$.]{
     \includegraphics[width=0.54\textwidth]{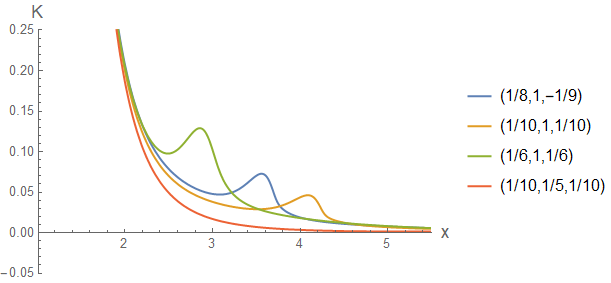}
           \label{KM4}} 
     \hfill
     \subfloat[A plot of the Kretschmann Scalar for the case of plot \ref{asympM3}, 
         showing a divergence as $x \rightarrow 1.90$.  This point is where the solution for $N$ tends to zero.]{
     \includegraphics[width=0.41\textwidth]{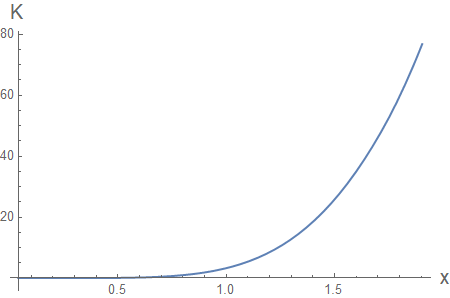}
           \label{KM3as}} 
        \caption{Plots of the curvature scalar for a horizon expansion solution and a solution for the asymptotic expansion.}
        \label{curv}
\end{figure*}

The plots below show the Kretschmann scalar for some special cases of the horizon and asymptotic expansions considered in Figs.\ \ref{horizonplots} and \ref{asymptoticplots}, along with that for the RN case.
These follow directly from the numerical code, 
and the same initial values are used.
Note that for the Kretschmann scalar in Fig.\ \ref{KM4}, 
the values of the potential strength $\t{g}$, the gravity coupling strength $k$, 
and the horizon value of $f'(1)$ are varied.
This plot shows the persistence of a bump feature indicating the rapid change of the metric functions shown in \ref{horizonM4}.

     % % \caption{ A plot of the Kretschmann scalar $K=R^{\mu\nu\ka\la}R_{\mu\nu\ka\la}$ for the case of plot \ref{horizonM4}, with small parameter variations.  The curvature near $x=1$ is actually finite (not shown).  We chose the vertical range to exhibit the interesting bump feature.  The curves are distinguished by the values of $(\t{g},k,f'(1))$.}
         %   \end{subfigure}
     % \begin{subfigure}[b]{0.42\textwidth}
     %     \centering
     %     \includegraphics[width=\textwidth]{KM3as.png}
     %     \caption{A plot of the Kretschmann Scalar for the case of plot \ref{asympM3}, 
     %     showing a divergence as $x \rightarrow 1.90$.  This point is where the solution for $N$ tends to zero. 
     %     }
     %     \label{KM3as}
     % \end{subfigure}

\begin{figure}[h]
  \centering
  \includegraphics[width=0.48\textwidth]{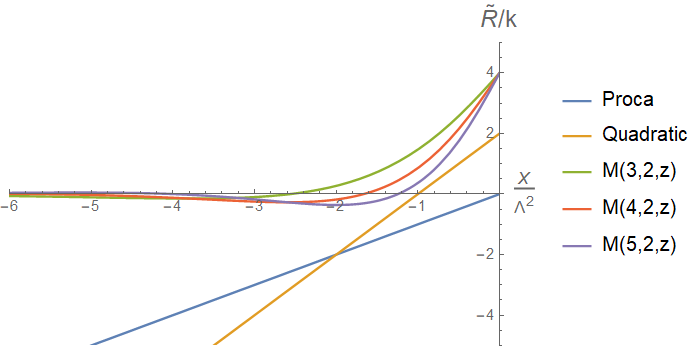}
         \caption{Plot of the (scaled) Ricci scalar for nonzero potential choices.  For the Einstein-Maxwell case $R=0$.  For the massive Proca potential and the quadratic potential, 
         the Ricci scalar diverges for decreasing $X / \La^2$.  For the hypergeometric potentials $R$ is bounded.}
         \label{ricci}
\end{figure}

Our aim here is not a complete study of curvature invariants. 
However, 
we close with an interesting analytical result that can be used to show an interesting feature of the hypergeometric potentials versus others.
Consider the Ricci scalar $R$ obtained by tracing the field equations \rf{FEgravBB}.
Due to the traceless nature of the field strength portion of the energy-momentum tensor, 
we find a general expression in terms of the potential only.
For any choice of $V(X)$, 
where $X=B^\mu B_\mu$, 
we find a remarkably simple result,
\beq
R=\ka (4V-2V'X).
\label{Rscalar}
\eeq
Note that $R=0$ for vanishing potential $V$, 
as expected.
For any hypergeometric potential the result \rf{Rscalar} is bounded, 
thus cannot diverge.
For the massive potential and quadratic potential it is not bounded.
The plot in Fig.\ \ref{ricci} demonstrates this.

\subsection{Asymptotic nonanalytic solution}
\label{non-analytic}

We found sinusoidal solutions for $B_0\sim \sin kr/r$, 
which are not analytic in $1/r$, 
in Section \ref{solutionsflat} on flat spacetime solutions. 
We examine here the possibility of these types of solutions in the context of the spherically symmetric coupled equations \rf{firstorder} or \rf{EFeqnsdim}.
As before we construct approximate solutions for the asymptotic region and use them to generate initial values for the full numerical solutions to the equations \rf{ueqns}.
The goal is to see if we get (i) oscillatory solutions and
(ii) solutions with horizons.

First we linearize the equations \rf{ueqns} around asymptotic values (values at $u=0$),
\beq
\bal
f &= f_0 + \ep \de f,\\
M &= M_0 + \ep \de M,\\
N &= N_0 + \ep \de N,
\label{expNA}
\eal
\eeq
where the asymptotic values are indicated with a subscript $0$, 
and $\ep$ is a small parameter to indicate a perturbative expansion.
The argument of $V$ and $V'$
is also expanded around an asymptotic value
$-\t{\be} = -f_0^2/N_0$, 
that minimizes the potential so that $V'(-\t{\be})=0$.  Note that $\t{\be}=\be/\La^2$, 
where $\be$ was defined in Sec.\ \ref{special solutions}.
We also take the minimum value of the potential $\bar V$ to be zero.
Upon inserting the expansions in equations \rf{expNA} into \rf{ueqns},
and keeping first order in $\ep$ terms, 
we obtain the following linear equations for the perturbations:
\beq
\bal
u^4 \de f'' &= b ( 2 N_0 \de f - f_0 \de N) \\
u^3 \de M' &= \ga b ( 2 N_0 \de f - f_0 \de N),\\
u \, \de N' &= \de N -2 \sqrt{N_0} \de M \\
\label{pertNA}.
\eal
\eeq
For consistency in the expansion we must have $M_0^2=N_0$.
The constants appearing are $b=-2 \t{\be} \tV^{\prime \prime}_0/N_0$ and 
$\ga= \mp k \sqrt{\t{\be}}/2$, with the plus or minus sign corresponding to the choice of minimum for $f_0=\pm \sqrt{\t{\be} N_0}$. 

Solutions to these equations can be found using standard methods by solving the coupled system \rf{pertNA}.
We find the following general solutions, 
neglecting extra constant terms,
\beq
\bal
\de f &= c_1 u e^{\sqrt{\rh}/u}  + \frac{c_2}{2\sqrt{\rh}} u e^{-\sqrt{\rh}/u}
\pm \frac{d_2 \sqrt{\t{\be}}}{\sqrt{N_0}(2-k\t{\be}) } u, \\
\de N &= \pm k \sqrt{\t{\be} N_0} \left( c_1 u e^{\sqrt{\rh}/u}  + \frac{c_2}{2\sqrt{\rh}} u e^{-\sqrt{\rh}/u}\right) 
\\
& \pt{= }
+ \frac{2 d_2}{2-k\t{\be} } u,\\
\de M &= \pm \frac {1}{2} k\sqrt{\t{\be}\rh}
\left( c_1 e^{\sqrt{\rh}/u}  - \frac{c_2}{2\sqrt{\rh}} e^{-\sqrt{\rh}/u}\right),
\label{pertNA2}
\eal
\eeq
where $\rh=(k\t{\be} -2) 2\t{\be}  \tV^{\prime \prime}_0$
and $c_1$, $c_2$, and $d_2$ are constants.
Note the interesting dependence on the minimum value of the vector field (scaled) $\t{\be}$.  We can have both oscillatory behavior or growing/decaying exponential behavior, 
similar to Ref.\ \cite{Bailey_2023}.

\begin{figure*}
    \centering
     % \begin{subfigure}[b]{0.49\textwidth}
       % \centering
    \subfloat[The quadratic potential case with initial data from \rf{pertNA}.  
   This shows initial oscillatory behavior followed by a naked singularity]{
         \includegraphics[width=0.48\textwidth]{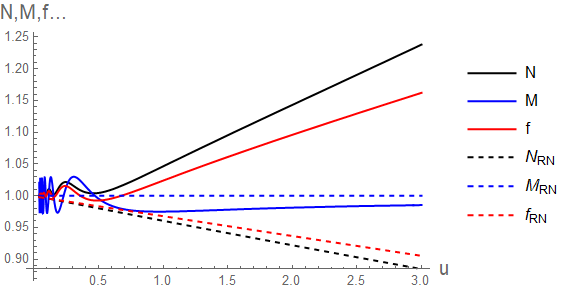}
         % \caption{The quadratic potential case with initial data from \rf{pertNA}.  This shows initial oscillatory behavior followed by a naked singularity.}
         \label{naQ}}
     % \end{subfigure}
    \hfill
     % \begin{subfigure}[b]{0.49\textwidth}
     %     \centering
     \subfloat[Close-up portion showing the contrast of the numerical solutions 
     with the solution from \rf{pertNA}.]{         
     \includegraphics[width=0.48\textwidth]{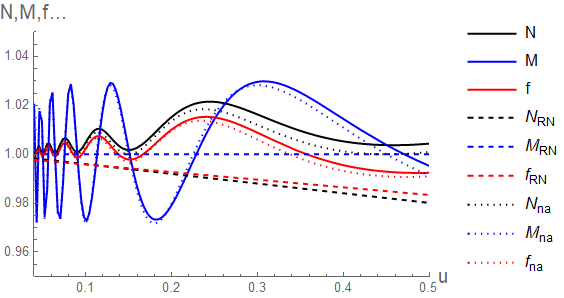}
         % \caption{Close-up portion showing the contrast of the numerical solutions with the solution from \rf{pertNA}.}
         \label{naQ2}}
         \\
     % \end{subfigure}
     % \begin{subfigure}[b]{0.49\textwidth}
     %     \centering
    \subfloat[The hypergeometric potential $M(3,2,z)$ case.]{      
    \includegraphics[width=0.48\textwidth]{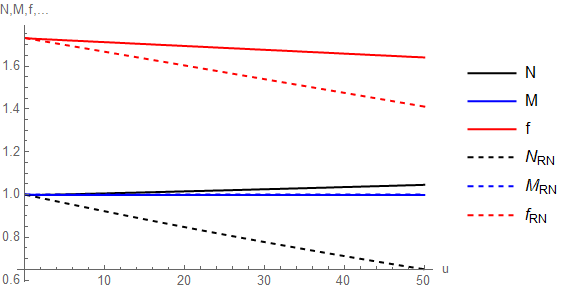}
         % \caption{The hypergeometric potential $M(3,2,z)$ case.}
         \label{naM3}}
     % \end{subfigure}
    \hfill
     % \begin{subfigure}[b]{0.49\textwidth}
     %     \centering
    \subfloat[The hypergeometric potential $M(3,2,z)$ case again;
         close-up portion of the left plot showing the contrast 
         of the numerical solutions with the perturbation solution from \rf{pertNA2}.]{        
             \includegraphics[width=0.48\textwidth]{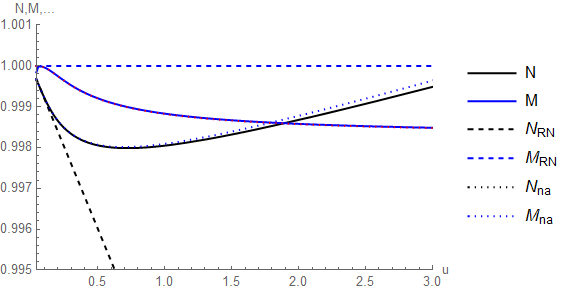}
         % \caption{The hypergeometric potential $M(3,2,z)$ case again;
         % close-up portion of the left plot showing the contrast of the numerical solutions with the perturbation solution from \rf{pertNA2}.}
         \label{naM32}}
     % \end{subfigure}
        \caption{Numerical solutions using the nonanalytic expansion \rf{pertNA} to generate initial values near $u=0$.  The results \rf{pertNA2} are labeled with subscript $na$.  
        The solid curves are the numerical code, 
        the dashed lines are the RN case, 
        and the dotted lines show the approximate solutions \rf{pertNA2}.
        We use the values $N_0=1=M_0$, with $\t{\be}=1$ (quadratic) and $\t{\be}=3$ (hypergeometric).
        We choose the constants to be ``small" $c_1 \sim 1/50$ for the quadratic case, 
        and $c_2 \sim 1/250$ for the hypergeometric case, to make the perturbation solution approximately valid.
        The connection with conventional Schwarzschild coordinates is $g_{tt}=A=-N$.}
        \label{NAasymptoticplots}
\end{figure*}

We examine the full numerical solution ``seeded" by these linearized approximations in order to determine how solutions that are not analytic at $u=0$ evolve with $u$.
We confine ourselves to two examples, for brevity, and we set the constants $\t{\la}$ and $\t{g}$ to unity.  
One is the hypergeometric potential with $n=3$ and other is the quadratic potential.
In Fig.\ \ref{NAasymptoticplots}, 
are two cases where $\t{\be}=3$ and $\t{\be}=1$, 
so that we capture damped behavior versus oscillatory.
For the former case we eliminate the terms growing with small $u$, with the positive exponential argument.
For the latter case, 
$c_2$ is related to $c_1$ such that the solutions are real-valued.
We use the same numerical scheme as described in Section \ref{asymptotic}, 
and input the initial values of $f$, $f'$, $N$ and $M$ at small u values $u=1/20$.

The plots \ref{naQ} and \ref{naM3} show the overall behavior while \ref{naQ2} and \ref{naM32} show behavior near $u=0$.
Both cases show agreement between the numerical solution and the approximation \rf{pertNA} for small $u$. 
In the case of the quadratic potential, 
the numerical solution clearly shows the same, 
small $u$, 
oscillatory behavior.
For the hypergeometric case, 
the approximate solution is not oscillatory because $\be=3$ so $\rh>0$, 
but it tracks the numerical solution closely for small $u$.
However, 
it appears both solutions have diverging $N$ as $u \rightarrow \infty$, 
indicating a naked singularity. 

\section{Discussion and Summary}
\label{discussion and summary}

\subsection{Observational constraints}
\label{observational constraints}

This work is preliminary in style.
The complications of solving nonlinear differential equations have been made evident in the preceding sections.
Since we do not have a general analytical solution for the cases where $V'\neq 0$, 
rather several classes of numerical solutions, 
it is challenging to test the parameters of the model in a systematic way.  In future work, 
this could be remedied.

However, 
we discuss a general strategy that could be used to measure parameters in this version of the bumblebee model, 
that involves orbital studies outside of the horizon where a post-Newtonian approach could be adopted.
Generally speaking, 
the effects of a black hole, 
apart from direct measurements of gravitational waves from black hole binary systems, 
are observed through its effects on orbiting bodies, including light.
As an example, consider the precision observations of high speed stars around the central black hole of the Milky Way \cite{Ghez:1998ph}.
The post-Newtonian acceleration for an orbiting body at radius $r$ takes the form: 
\beq
\vec a = - \frac {GM}{r^2} \hat r + \de \vec a, 
\label{accel}
\eeq
where $ \de \vec a$ contains all relevant effects (e.g., other nearby bodies, relativistic effects, beyond GR effects, in a post-Newtonian series, etc. as in standard references like Ref.\  \cite{Will:2018bme}).
Indeed, 
this type of equation has been used for orbital studies in a plethora of works to constrain local Lorentz violation in gravity before (e.g., Refs.\ \cite{bk05,Bourgoin:2016ynf,Shao:2014oha, Shao:2015gua, Hees:2015mga}).
Here we are interested in the form of perturbations $\de \vec a$ stemming from the solutions found in this work in Section \ref{gravity-vector part II}.

\begin{figure}[h]
  \centering
         \includegraphics[width=0.50\textwidth]{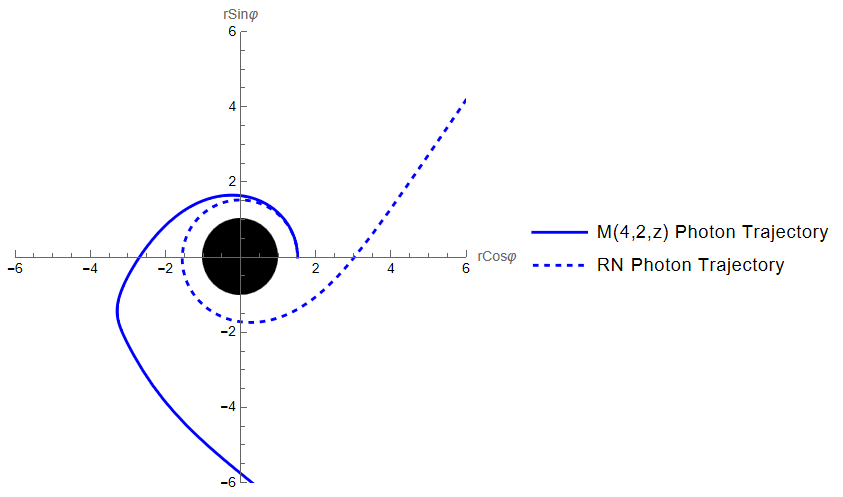}
         \caption{Sample null geodesic trajectories of the RN metric and the $M(4,2,z)$ metric of Fig.\ \ref{horizonM4}.  The black circle is the horizon and the axes are the spherical projections onto the orbital plane with $\th=\pi/2$.}
         \label{orbit}
\end{figure}

Consider the case of the solution in Fig.\ \ref{horizonM4}, 
the horizon expansion, 
for the hypergeometric potential $V=g\La^4 (M(4,2,z)+ 1/3e^2)$ (which vanishes at the minimum value $z=X/\La^2=-2$).
This solution has a horizon, 
by construction, 
but also has two smoking gun features.
One is the peculiar peak for $M$, and resulting ``dip" in $N$, 
and the other is the steady decrease to zero afterwards, 
both in stark contrast to the standard behavior of the RN solution.
Some preliminary work on plotting orbits shows the effect of such a metric.
In Fig.\ \ref{orbit}, 
we show null geodesic trajectories of the RN metric and the $M(4,2,z)$ metric of Fig.\ \ref{horizonM4}, 
with the same initial values of position and velocity.
One can see the exaggerated features of the orbit in the symmetry-breaking case (a hypergeometric potential).  
We leave a thorough analysis of orbits for future work.

Following up on this example, 
one could crudely estimate $\de \vec a$ from examining the weak field geodesic equation for the radial acceleration, $a^r \approx -\Ga^r_{\pt{r}tt} \approx \tfrac 12 \prt_r g_{tt}$.
In terms of $N(x)$, 
this would be $a^r = - (1/2r_h)dN/dx $.
Measurements of accelerations of objects around black holes can yield results for orbital quantities indicating errors at the $1\%$ level or less \cite{Ghez:2003qj}.  
Thus, 
we could expect to place limits on acceleration perturbations at this level.
What is needed, 
for the horizon expansion solutions, 
is a numerical study of the dependence of $dN/dx$ on the values of three relevant parameters $g$ and $k$, and the initial value $\af1$.
One could then use data to extract a statistical distribution for these parameters.

Some study of the variation of the three parameters for this case shows that for $k\approx 3 \times 10^{-5}$, 
$\af1=1/10$, 
and $\t{g}=1/100$ results in a $\De a/a \sim 1/100$
where $\De a$ is the difference in the radial accelerations between the RN solution and the solution with the $M(4,2,z)$ potential case.  
Note that these are not remarkably small values, 
as is typically suggested for spacetime symmetry breaking \cite{kp95}.
Indeed, previous studies have shown that large Lorentz violation is possible in some sectors even with present experimental limits \cite{kt09,bkx15,Bailey_2023,Kostelecky:2024rsn}.

For the asymptotic expansion approach of Section \ref{asymptotic}, 
similar approaches as described for the horizon expansion could be adopted.
That is, 
one explores the effective acceleration of a test body numerically, 
as a function of the parameters $\t{g}$, $k$, 
and in this case $\cf0$ and $\cf1$.

\subsection{Summary}
\label{summary}

In this work, 
we systematically studied
the case of the bumblebee gravity model contained in the Lagrange density \rf{bbmodel}, 
subject to vanishing nonminimal couplings \rf{xi0}.
The main goal in this paper is exploring spherically symmetric, static solutions for the case when the field configuration for $B_\mu$ does not lie at the minimum of the potential ($V' \neq 0$).

In flat spacetime,
Sec.\ \ref{flat spacetime study},
we established some general results for 
the spherical symmetry case.
One key result is the analysis of the Hamiltonian for a generic potential $V(X)$, 
where $X=B_\al B^\al$, Eq.\ 
\rf{hamOS}.
We discussed properties of hypergeometric potentials \rf{hyper} 
and contrasted them with polynomial potentials.
For example, 
see the plot of the Hamiltonian density for the case $\vec B=0$,
shown in Fig.\ \ref{potsplot}.

Also in flat spacetime, 
we discussed the solution to the bumblebee field equation in the spherical symmetry case in Section \ref{spherical nonlinear}.
The perturbative solution around the minimum $b_\mu$
was obtained in equation \rf{linsolns}.
For the fully nonlinear case, 
we characterized the asymptotic behavior of the solutions for different choices of initial conditions in equations \rf{Quadratic Stability Argument}-\rf{Hypergeometric 4 Stability Argument}, 
which can be seen also in Figs.\ \ref{Quadratic Stability pdf}-\ref{n4 Hyper Stability pdf}.

In Section \ref{gravity-vector, part I}, 
we investigated some general aspects of the field equations with gravity.
First, 
we briefly discussed energy conditions, 
followed by a sample solution showing that AdS spacetime could arise in some cases in \rf{friedman}.
Then, 
in Sec.\ \ref{bekenstein}
we examined the Bekenstein horizon argument, 
resulting in equation \rf{int3},
and concluded there are no constraints on the bumblebee model with arbitrary potentials.
Another important set of results are the special case solutions, 
where analytical solutions are possible in Sec.\ \ref{special solutions}.
The chart in Fig.\ \ref{flowchart} categorized these solutions.

Numerical solutions to the full equations were the subject of Sec. \ref{gravity-vector part II}.
Firstly, 
we established the field equations in Eddington-Finkelstein coordinates
in equations \rf{EFeqns} and \rf{EFeqnsdim}.
In Sec.\ \ref{analytic}, 
we then focused on the horizon region and the asymptotic regions, 
using two different series expansions \rf{horizonexp} and \rf{asymptoticexp}, 
with lower-order coefficients in Tables \ref{horizon1}-\ref{asymptotic2}.
These expansions were then used to seed a full numerical solver.
Sample solutions for different choices of boundary values are shown 
in Fig.\ \ref{horizonplots} for the horizon expansion in Fig.\ \ref{asymptoticplots} for the asymptotic expansion.

We primarily studied four types of potentials:
Proca (massive vector), quadratic, and two 
hypergeometric choices $M(n,2,z)$ for $n=3$ and $n=4$.
General features observed include sensitivity to initial values at the horizon $r=r_h$ or at infinity $x \rightarrow \infty$.
We found naked singularities as possibilities, as illustrated in Fig. \ref{asympQ2}.
Also we found wildly varying behavior of the metric functions (equivalent to changing signs of $\prt_r g_{tt}$ some distance from the horizon), 
as illustrated in Fig. \ref{horizonM4}.
In addition, 
the curvature evaluated at the horizons or at infinity, 
was found to diverge, 
or converge, 
depending on the type of potential, 
and the initial values chosen.
The latter can be seen in Figs.\ \ref{KM4} and \ref{KM3as}.

In Sec.\ \ref{non-analytic},
we also studied the nonanalytic perturbative expansions of the field equations around the background values.
The solutions are contained in Eq.\ \rf{pertNA2}.
These were again used to seed the numerical solutions.
The results showed oscillatory and naked singularity behavior as seen in Fig.\ \ref{NAasymptoticplots}.
In Sec.\ \ref{observational constraints}, 
we discussed the effects on  trajectories of orbits in the vicinity of the black hole solutions, 
and how the model parameters might be constrained by observation.

We view this work as a ``door opening" into the rich structure of possibilities of solutions with $V' \neq 0$.
Since we showed that such models with hypergeometric potential functions do not automatically result in energies unbounded from below, 
it represents a complimentary area to explore, 
alongside literature on the couplings $\xi_1$ and $\xi_2$.
It would be of interest to study these types of solutions further.
As already outlined above, 
constructing a numerical scheme that varies the parameters in the model,
while calculating an observable, 
such as an orbital effect,
and then comparing against orbital data would be of interest.
This may involve a step up in computing time, 
given that the numerical solver used in this work was at least mildly time-consuming on an ordinary computer.

Another aspect that should be studied is the stability of the analytical (RN-like and AdS/dS-like) solutions derived in Section \ref{special solutions}.
For instance, 
one could look for perturbations of the metric and bumblebee field around these solutions to check stability properties \cite{Mai:2024lgk}.
Other items left unexplored include allowing for generic $B_r \neq 0$ components, the spacelike $B_\mu$ case, 
and also looking for Kerr-like solutions in an axisymmetric spacetime.  
Of course one can also study cases including not only the $V' \neq 0$, but also nonzero values for the nonminimal couplings in \rf{bbmodel}.
Future exploration may show that the bumblebee still can fly.

\begin{acknowledgments}
The authors thank Nils A.\ Nilsson, Robertus Potting, and Edward Poon for useful discussions that improved the manuscript.
For Q.G.B., H.S.M. and D.W.-C.\ this work was supported by the National Science Foundation grant number 2308602.
H.S.M.\ and D.W.-C.\ were supported by the Embry-Riddle Aeronautical University Undergraduate Research Institute and the Arizona NASA Space Grant consortium.
\end{acknowledgments}

\bibliography{refsv3}

\end{document}